\documentclass[10pt]{article}
\usepackage{fullpage}
\usepackage{amsmath}
\usepackage{amssymb}
\usepackage[dvips]{epsfig}
\usepackage{color}

\def\##1{{\bf #1}}
\def\=#1{\underline{\underline #1}}

\def\eps{\epsilon}
\def\epso{\epsilon_0}
\def\muo{\mu_0}
\def\ko{k_0}
\def\co{c_0}

\def\etao{\eta_0}
\def\.{\mbox{ \tiny{$^\bullet$} }}

\def\le{\left(}
\def\ri{\right)}
\def\les{\left[}
\def\ris{\right]}
\def\lec{\left\{}
\def\ric{\right\}}

\def\c#1{\cite{#1}}
\def\l#1{\label{#1}}
\def\r#1{(\ref{#1})}

\begin{document}
\begin{center}

{\bf {\LARGE Surface--plasmon--polariton waves
 guided by the uniformly moving planar interface of a metal film and
dielectric slab }}

\vspace{10mm} \large

 Tom G. Mackay\footnote{E--mail: T.Mackay@ed.ac.uk.}\\
{\em School of Mathematics and
   Maxwell Institute for Mathematical Sciences\\
University of Edinburgh, Edinburgh EH9 3JZ, UK}\\
and\\
 {\em NanoMM~---~Nanoengineered Metamaterials Group\\ Department of Engineering Science and Mechanics\\
Pennsylvania State University, University Park, PA 16802--6812,
USA}\\
 \vspace{3mm}
 Akhlesh  Lakhtakia\footnote{E--mail: akhlesh@psu.edu}\\
 {\em NanoMM~---~Nanoengineered Metamaterials Group\\ Department of Engineering Science and Mechanics\\
Pennsylvania State University, University Park, PA 16802--6812, USA}

\normalsize

\vspace{15mm} {\bf Abstract}

\end{center}

\vspace{4mm}

We explored the effects of relative motion on the excitation of
surface--plasmon--polariton (SPP) waves guided by the planar
interface of a metal film and a dielectric slab, both materials
being isotropic and homogeneous. Electromagnetic phasors in moving
and non--moving reference frames were related directly using the
corresponding Lorentz transformations. Our numerical studies
revealed that, in the case of a uniformly moving dielectric slab,
the angle of incidence for SPP-wave excitation is highly sensitive
to
 (i) the ratio $\beta$ of the speed of motion to speed of light in
free space and (ii) the direction of motion. When the
direction of motion is
 parallel to the
plane of incidence, the SPP wave is excited by $p$-polarized (but
not $s$-polarized) incident plane waves for low and moderate values
of $\beta$, while at higher values of $\beta$ the total reflection
regime breaks down. When the direction of motion is
 perpendicular to the
plane of incidence, the SPP wave is excited by $p$-polarized
incident plane waves for low values of $\beta$, but $s$-polarized
incident plane waves at moderate values of $\beta$,  while at higher
values of $\beta$ the SPP wave is not excited.
 In
the case of a uniformly moving metal film, the
 sensitivity  to $\beta$ and the direction of motion is less obvious.

 \vspace{5mm}


\section{Introduction}

In  quantum mechanical terms,  a surface plasmon--polariton (SPP) is
a quasiparticle. Created by the interaction of photons in a
dielectric material and electrons in a metal, a SPP travels along
the interface of the metal and the dielectric material \c{Felbacq}.
In classical terms, a SPP wave propagates guided by the interface
 with an amplitude that decreases exponentially with distance from the interface.
Over the past few decades, SPP waves  have been widely investigated,
in part because of the opportunities that they present for optical
sensing applications \c{Homola2008,AZL2}. Most published research
deals with SPP waves guided by the interface of a metal with an
isotropic dielectric material, both homogeneous
\c{Kret,Agran,PLreview}.  Guidance by the interface of a metal with
an anisotropic dielectric material has also been examined
\c{Sprokel1,Sprokel2,Lloyd,Singh,Mihalache,Kano}.
 The interface of a metal with a periodically
nonhomogeneous and anisotropic dielectric material has recently been
found to support more than one mode of SPP-wave propagation at a
fixed frequency \c{Polo_PRSA,Motyka1,Motyka_II}.

In this communication, we address a fundamental issue: the effect of
relative uniform motion  on SPP-wave propagation guided by the
planar interface of a metal and a dielectric material, both
isotropic and homogeneous. This is done using a modification
\c{Lakh_oc} of the standard Kretschmann configuration \c{Kret}, as
detailed in Sec.~\ref{analysis}. Electromagnetic phasors in moving
and non--moving (laboratory) inertial reference frames are related
directly using the corresponding Lorentz transformations
\c{BC1967,ML2009}, and not the Minkowski constitutive relations
\cite[Chap.~8]{Chen}. Section~\ref{ns} provides numerical results,
and some remarks in Sec.~\ref{cr} close this communication.

As regards notation, vectors and matrixes are written in boldface,
with the $\hat{}$ symbol denoting a unit vector. Square brackets
enclose 2--, 4-- and 6--vectors, as well as 2$\times$2,  4$\times$4
and 6$\times$6 matrixes.  Double underlining indicates a 3$\times$3
dyadic, with the identity dyadic $\=I = \hat{\#x}\,\hat{\#x} +
\hat{\#y}\,\hat{\#y} + \hat{\#z}\,\hat{\#z}$ and the null dyadic
being $\=0$. The permittivity and permeability of free space are
denoted by $\epso$ and $\muo$;  the free--space wavenumber at
angular frequency $\omega$ is $\ko = \omega \sqrt{\epso \muo}$; and
$\co = \omega / \ko$.

\section{Analysis} \l{analysis}

The partition of space into four distinct layers provides the
backdrop for our analysis:
\begin{itemize}
\item[(i)] the half--space $z< 0$ is occupied by
a  dielectric material with  relative permittivity scalar $\eps_i$ with
respect to the inertial reference frame $\Pi$, which is the
laboratory frame;
\item[(ii)] a thin metal film of
relative permittivity scalar $\eps_m$ with respect to the inertial reference
frame $\Pi$ fills the layer $0 < z < L_m$;
\item[(iii)] a dielectric slab
of   relative permittivity scalar $\eps'_d$ with respect to the inertial
reference frame $\Pi'$ fills the layer $L_m < z < L_\Sigma$; and
\item[(iv)] the half--space $z > L_\Sigma$ is occupied by a  dielectric
material with relative permittivity scalar $\eps_t$ with respect to the
inertial reference frame $\Pi$.
\end{itemize}
All four materials are homogeneous and
their relative permittivity scalars are frequency-dependent in their respective co-moving inertial reference
frames. Dissipation is small enough to be ignored in both materials occupying the two half--spaces
as well as in the dielectric slab.
The reference  frame $\Pi'$ moves at
uniform velocity $\#v = v \hat{\#v}$, in the $xy$ plane, with respect to the laboratory frame $\Pi$.
  This setup, schematically
illustrated in Fig.~\ref{fig1} for the case $\hat{\#v} = \hat{\#x}$,
represents a modification \c{Lakh_oc} of the standard Kretschmann
configuration \c{Kret}, suitable for launching SPP waves guided by the
planar interface of the metal film and the dielectric slab; i.e.,
along $z=L_m$.

Suppose that an arbitrarily polarized plane wave in the half--space
$z < 0$ is directed towards the metal film. Its wavevector lies in
the $xz$ plane, making an angle $\theta_{i} \in \les 0, \pi/2 \ri$
to the $+z$ axis, in the laboratory frame $\Pi$. In consequence,
a reflected plane wave is generated in the half--space $z < 0$ along
with a transmitted plane wave in the half--space $z
> L_\Sigma$. In the laboratory frame,
the total electric field phasor in the half--space $z < 0$ may be
written as
\begin{eqnarray}
  \#E (\#r, \omega) & =&   \big\{ \les \,
a_s \#u_y + a_p \#p_+ (\theta_{i})  \ris \exp \le i \ko \sqrt{\eps_i
} \,  z \cos \theta_{i} \ri  + \les \, r_s \#u_y + r_p \#p_-
(\theta_{i}) \ris  \exp \le - i \ko \sqrt{\eps_i } \,  z \cos
\theta_{i} \ri \big\} \nonumber
\\ &&  \times \exp \le i \kappa x \ri, \qquad
 z < 0,
\end{eqnarray}
whereas that in the half--space $z > L_\Sigma$ may be written as
\begin{eqnarray}
&& \#E (\#r, \omega) = \les \, t_s \#u_y + t_p \#p_+ (\theta_{t})
\,\ris \exp \les i \ko \sqrt{\eps_t } \le z - L_\Sigma \ri \cos
\theta_{t} \ris \, \exp \le i \kappa x \ri ,
 \qquad z > L_\Sigma.
\end{eqnarray}
Here $\#p_\pm (\theta)  = \mp \#u_x \cos \theta + \#u_z \sin
\theta$, $\kappa=\ko \sqrt{\eps_i} \, \sin \theta_{i}$ and the angle
of transmission $\theta_{t}$ in frame $\Pi$ satisfies
\begin{equation}
\sqrt{\eps_{i}} \, \sin \theta_{i} = \sqrt{\eps_{t}} \, \sin
\theta_{t}.
\end{equation}
Our task is to relate the complex--valued reflection and
transmission amplitudes---namely $r_{s}$, $r_{p}$, $t_{s}$ and
 $t_{p}$---to the  corresponding amplitudes $a_s$ and $a_p$ of the $s$- and
$p$-polarized components of the incident plane wave.

For later use,
we note that the wavevector
\begin{equation} \#k_i = \kappa \, \hat{\#x} + \ko \sqrt{\eps_i} \,
\cos \theta_i \, \hat{\#z}, \qquad z< 0,
\end{equation}
of the incident plane wave in frame
$\Pi$
is related to its counterpart $\#k'_i$ in frame $\Pi'$ by
the Lorentz transformation \c{Chen,EAB}
\begin{equation}
 \#k_{i} = \displaystyle{ \gamma \le \#k'_{i} \cdot \hat{\#v} + \frac{\omega' v}{\co^2}
\ri \hat{\#v} + \le \, \=I - \hat{\#v} \, \hat{\#v} \ri \cdot
\#k'_{i}}, \l{k_LT}
\end{equation}
where the scalar parameters
\begin{equation}
 \gamma = \frac{1}{\sqrt{1-\beta^2}}, \qquad \beta = \frac{v}{\co}.
 \end{equation}
The angular frequencies in the two frames are related per
\begin{equation}
 \omega = \displaystyle{\gamma \le \omega' + \#k'_{i} \cdot \#v
 \ri}.
 \l{o_LT}
\end{equation}

In the metal film, the electric and magnetic field phasors with
respect to the laboratory frame $\Pi$, namely $\#E (\#r, \omega) = \#e(z, \kappa,
\omega)\, \exp \le i \kappa x \ri $ and $\#H (\#r, \omega) = \#h(z,
\kappa, \omega) \, \exp \le i \kappa x \ri $, are related via the
Maxwell curl postulates; thus,
\begin{eqnarray} \l{MODE_m}
\frac{d}{d z} \les \#f (z, \kappa, \omega )\ris &=& i \les \#P
(\eps_m, \kappa, \omega ) \ris \, \les \#f (z, \kappa, \omega )\ris,
\qquad 0 < z < L_m,
\end{eqnarray}
where the 4--vector
\begin{equation}
\les \#f (z, \kappa, \omega ) \ris = \les \begin{array}{c} \#e(z,
\kappa, \omega) \cdot \hat{\#x} \\
\#e(z, \kappa, \omega) \cdot \hat{\#y} \\ \#h(z,
\kappa, \omega) \cdot \hat{\#x} \\
\#h(z, \kappa, \omega) \cdot \hat{\#y} \end{array} \ris
\end{equation}
and the 4$\times$4 matrix
\begin{eqnarray}
\hspace{-20pt} \les  \#P (\eps, \kappa, \omega ) \ris = \omega \les
\begin{array}{cccc}
0 &0 & 0 &  \muo - \frac{ \kappa^2}{\omega^2 \epso \eps
} \\
0 &0 & -  \muo &0 \\
0 & \frac{ \kappa^2}{\omega^2 \muo } -  \epso \eps  & 0
&0 \\
 \epso \eps  &0&0&0
\end{array} \ris. \nonumber \\ &&
\end{eqnarray}
The solution to the matrix ordinary differential equation \r{MODE_m}
is conveniently stated as
\begin{equation}
\les \#f (L_m, \kappa, \omega )\ris = \les \#M (\eps_m, L_m, \kappa,
\omega )\ris \les \#f (0, \kappa, \omega )\ris, \l{sol_m}
\end{equation}
where the transfer matrix
\begin{equation}
\les \#M (\eps_m, L_m, \kappa, \omega )\ris = \exp \lec i L_m  \les
\#P (\eps_m,  \kappa, \omega )\ris  \ric.
\end{equation}

In a similar fashion,  the electric and magnetic field phasors with
respect to frame $\Pi'$, namely $\#E' (\#r', \omega') = \#e'(z,
\kappa', \omega')\, \exp \le i \kappa' x' \ri $ and $\#H' (\#r',
\omega') = \#h'(z, \kappa', \omega') \, \exp \le i \kappa' x' \ri $,
satisfy the matrix ordinary differential equation
\begin{eqnarray} \l{MODE_d}
\frac{d}{d z} \les \#f' (z, \kappa', \omega' )\ris &=& i \les \#P
(\eps'_d, \kappa', \omega' ) \ris \, \les \#f' (z, \kappa', \omega'
)\ris, \qquad L_m < z < L_\Sigma
\end{eqnarray}
in the dielectric slab, where the 4--vector
\begin{equation}
\les \#f' (z, \kappa', \omega' ) \ris = \les \begin{array}{c}
\#e'(z,
\kappa', \omega') \cdot \hat{\#x} \\
\#e'(z, \kappa', \omega') \cdot \hat{\#y} \\ \#h'(z,
\kappa', \omega') \cdot \hat{\#x} \\
\#h'(z, \kappa', \omega') \cdot \hat{\#y} \end{array} \ris,
\end{equation}
 the angular frequency $\omega'$ is specified by \r{o_LT} and
$\kappa' = \#k'_i \cdot \hat{\#x}$ with $\#k'_i$ being specified by
\r{k_LT}.  The
 solution to the matrix ordinary differential equation \r{MODE_d}
is conveniently stated as
\begin{equation}
\les \#f' (L_\Sigma, \kappa', \omega' )\ris = \les \#M (\eps'_d,
L_\Sigma-L_m, \kappa', \omega' )\ris \les \#f' (L_m, \kappa',
\omega' )\ris. \l{soln_d}
\end{equation}

We now seek to express the solution  \r{soln_d} in terms of the field
phasors in the laboratory frame
$\Pi$. Using the fact that the $z$ components of $\#e'(z,
\kappa', \omega')$ and $\#h'(z, \kappa', \omega')$ are related via
the Maxwell curl postulates per
\begin{equation}
\left.
\begin{array}{l}
\#e'(z, \kappa', \omega') \cdot \hat{\#z} = -
\displaystyle{\frac{\kappa'}{\omega' \epso \eps'_d} \, \#h'(z,
\kappa', \omega') \cdot \hat{\#y}} \vspace{4pt} \\
\#h'(z, \kappa', \omega') \cdot \hat{\#z} =
\displaystyle{\frac{\kappa'}{\omega' \muo} \, \#e'(z, \kappa',
\omega') \cdot \hat{\#y}}
\end{array}
\right\},
\end{equation}
we introduce the 6--vector extension of $\les \#f' (z, \kappa',
\omega' ) \ris$, namely
\begin{equation}
\les \tilde{\#f}' (z, \kappa', \omega' ) \ris = \les
\begin{array}{c} \#e'(z,
\kappa', \omega') \cdot \hat{\#x} \\
\#e'(z, \kappa', \omega') \cdot \hat{\#y}
 \\
  \#e'(z,
\kappa', \omega') \cdot \hat{\#z} \\ \#h'(z,
\kappa', \omega') \cdot \hat{\#x} \\
\#h'(z, \kappa', \omega') \cdot \hat{\#y} \\
 \#h'(z,
\kappa', \omega') \cdot \hat{\#z}  \end{array} \ris,
\end{equation}
and the 6$\times$6 matrix extension of $ \les \#M (\eps'_d,
L_\Sigma-L_m, \kappa', \omega' )\ris $, namely
$
 \les \tilde{\#M} (\eps'_d,
L_\Sigma-L_m, \kappa', \omega' )\ris $, with components
\begin{eqnarray}
&& \les \tilde{\#M} (\eps'_d, L_\Sigma-L_m, \kappa', \omega'
)\ris_{\tilde{p} \tilde{q}} =
 \les \#M (\eps'_d,
L_\Sigma-L_m, \kappa', \omega' )\ris_{p q }, \qquad \tilde{p},
\tilde{q} \in \lec 1, 2, 4, 5 \ric; \quad p, q \in \lec 1, 2, 3, 4
\ric ; \nonumber \\ && \\
&& \les \tilde{\#M} (\eps'_d, L_\Sigma-L_m, \kappa', \omega'
)\ris_{3 \tilde{q}} =  - \displaystyle{\frac{\kappa'}{\omega' \epso
\eps'_d} \,
 \les \#M (\eps'_d,
L_\Sigma-L_m, \kappa', \omega' )\ris_{4 q }, \qquad  \tilde{q} \in
\lec 1, 2, 4, 5 \ric; \quad  q \in \lec 1, 2, 3, 4 \ric ;} \nonumber
\\ && \\
&& \les \tilde{\#M} (\eps'_d, L_\Sigma-L_m, \kappa', \omega'
)\ris_{6 \tilde{q}} =   \displaystyle{\frac{\kappa'}{\omega' \muo }
\,
 \les \#M (\eps'_d,
L_\Sigma-L_m, \kappa', \omega' )\ris_{2 q }, \qquad  \tilde{q} \in
\lec 1, 2, 4, 5 \ric; \quad  q \in \lec 1, 2, 3, 4 \ric ;} \nonumber
\\ && \\
&& \les \tilde{\#M} (\eps'_d, L_\Sigma-L_m, \kappa', \omega'
)\ris_{\tilde{p} \tilde{q}} = 0, \qquad  \tilde{p}, \tilde{q} \in
\lec 3, 6 \ric.
\end{eqnarray}
Thus, the solution \r{soln_d} may be extended as
\begin{equation}
\les \tilde{\#f}' (L_\Sigma, \kappa', \omega' )\ris = \les
\tilde{\#M} (\eps'_d, L_\Sigma-L_m, \kappa', \omega' )\ris \les
\tilde{\#f}' (L_m, \kappa', \omega' )\ris. \l{soln_d2}
\end{equation}
Now in the dielectric slab the electromagnetic phasors in frames
$\Pi$ and $\Pi'$ are related by the Lorentz transformations \c{EAB}
\begin{equation}
\left.\begin{array}{l}
\#e (z, \kappa, \omega) = \=A_{\,d} \cdot \#e' (z, \kappa',
\omega')\\[5pt]
\#h (z, \kappa, \omega) = \=A_{\,d} \cdot \#h' (z, \kappa',
\omega')
\end{array}\right\}, \qquad L_m < z < L_\Sigma,
\end{equation}
where the 3$\times$3 dyadic
\begin{equation}
\=A_{\,d} = \le 1 - \gamma \ri \hat{\#v}\, \hat{\#v} + \gamma \=I -
\frac{\gamma}{\omega'} \le \#v \times  \=I \, \ri \cdot \le \#k'_d
\times \=I \, \ri,
\end{equation}
with
\begin{equation} \#k'_d =
\kappa'\hat{\#x}' + \sqrt{\le \omega' \ri^2 \epso \eps'_d \muo - \le
\kappa' \ri^2} \; \; \hat{\#z}, \qquad L_m < z < L_\Sigma.
\end{equation}
Thus, \r{soln_d2} leads to
\begin{equation}
\les \tilde{\#f} (L_\Sigma, \kappa, \omega )\ris = \les \tilde{\#N}
(\eps'_d, L_\Sigma-L_m, \kappa, \omega )\ris \les \tilde{\#f} (L_m,
\kappa, \omega )\ris \l{soln_d3}
\end{equation}
in the laboratory frame $\Pi$,
with the 6$\times$6 matrix
\begin{equation}
\les \tilde{\#N} (\eps'_d, L_\Sigma-L_m, \kappa, \omega )\ris  =
\les \begin{array}{cc} \=A_{\,d} & \=0 \vspace{4pt} \\
\=0 & \=A_{\,d} \end{array} \ris^{-1} \les \tilde{\#M} (\eps'_d,
L_\Sigma-L_m, \kappa', \omega' )\ris \les \begin{array}{cc} \=A_{\,d} & \=0 \vspace{4pt} \\
\=0 & \=A_{\,d} \end{array} \ris
\end{equation}
and the 6--vector
\begin{equation}
\les \tilde{\#f} (z, \kappa, \omega ) \ris = \les
\begin{array}{c} \#e(z,
\kappa, \omega) \cdot \hat{\#x} \\
\#e(z, \kappa, \omega) \cdot \hat{\#y}
 \\
  \#e(z,
\kappa, \omega) \cdot \hat{\#z} \\ \#h(z,
\kappa, \omega) \cdot \hat{\#x} \\
\#h(z, \kappa, \omega) \cdot \hat{\#y} \\
 \#h(z,
\kappa, \omega) \cdot \hat{\#z}  \end{array} \ris.
\end{equation}
Finally,  the counterpart of \r{soln_d} in the laboratory frame $\Pi$ emerges as
\begin{equation}
\les \#f (L_\Sigma, \kappa, \omega )\ris = \les \#N (\eps'_d,
L_\Sigma-L_m, \kappa, \omega )\ris \les \#f (L_m, \kappa, \omega
)\ris, \l{soln_d5}
\end{equation}
wherein the components of the 4$\times$4 matrix $ \les \#N (\eps'_d,
L_\Sigma-L_m, \kappa, \omega )\ris$ are given by
\begin{eqnarray}
&& \les \#N (\eps'_d, L_\Sigma-L_m, \kappa, \omega )\ris_{p q} =
 \les \tilde{\#N} (\eps'_d,
L_\Sigma-L_m, \kappa, \omega )\ris_{\tilde{p}  \tilde{q} }, \qquad
p, q \in \lec 1, 2, 3, 4 \ric ; \quad \tilde{p}, \tilde{q} \in \lec
1, 2, 4, 5 \ric.   \nonumber \\ &&
\end{eqnarray}

 For later use, we introduce the 4$\times$4 matrix $ \les \#Q
(\eps'_d,  \kappa, \omega )\ris$ via
\begin{equation}
 \les \#N (\eps'_d, L_\Sigma-L_m, \kappa, \omega )\ris = \exp \lec i
 \le L_\Sigma - L_m \ri  \les \#Q (\eps'_d,
 \kappa, \omega )\ris \ric.
\end{equation}
The matrixes $ \les \#N (\eps'_d, L_\Sigma-L_m, \kappa,
\omega )\ris$ and $\les \#Q (\eps'_d, \kappa, \omega )\ris$ have the same eigenvectors.
The eigenvalues $ \sigma_\ell$, $\ell \in \lec 1,2,3,4 \ric$, of   $ \les \#N (\eps'_d, L_\Sigma-L_m, \kappa,
\omega )\ris$  are related to the eigenvalues $\alpha_\ell$ of  $
\les \#Q (\eps'_d, \kappa, \omega )\ris$ as follows:
\begin{equation}
\alpha_\ell = \frac{\ln \sigma_\ell}{i \le L_\Sigma - L_m \ri},
\qquad \quad \ell \in \lec 1,2,3,4 \ric.
\end{equation}

Upon combining the solutions \r{sol_m} and \r{soln_d5},  and
invoking  the boundary conditions on the tangential components of
the electric and magnetic phasors at the planar interfaces, we
arrive at the algebraic relation
\begin{eqnarray}
\les
\begin{array}{c}
t_s \\ t_p \\ 0 \\ 0 \end{array} \ris \hspace{-4pt} & = &
\hspace{-4pt}\les \,\#K (\eps_t, \theta_{t}) \,\ris^{-1}  \les \#N
(\eps'_d, L_\Sigma-L_m, \kappa, \omega )\ris \les \#M (\eps_m, L_m,
\kappa, \omega )\ris  \les \,\#K (\eps_i, \theta_{i}) \,\ris  \les
\begin{array}{c}
a_s \\ a_p \\ r_s \\ r_p \end{array} \ris, \l{bvp}
\end{eqnarray}
where the 4$\times$4 matrix
\begin{eqnarray}
&& \hspace{-20pt}   \les \,\#K (\eps, \theta) \,\ris = \les
\begin{array}{cccc}
0 & - \cos \theta & 0 & \cos \theta \\ 1 &0 &1 & 0\\
- \frac{ \sqrt{\eps} \, \cos \theta }{\etao} &0&  \frac{\sqrt{\eps}
\, \cos \theta }{\etao} & 0
\\ 0 & - \frac{\sqrt{\eps }}{\etao} &0 & - \frac{\sqrt{\eps
}}{\etao}
\end{array} \ris. \nonumber \\ &&
\end{eqnarray}
A straightforward manipulation of \r{bvp} yields
\begin{equation}
\les \begin{array}{c} r_s \vspace{4pt} \\
r_p \end{array} \ris = \les \begin{array}{cc} r_{ss}  &
r_{sp} \vspace{4pt} \\ r_{ps} & r_{pp}
\end{array} \ris
 \les \begin{array}{c} a_s \vspace{4pt} \\
a_p \end{array} \ris
\end{equation}
and
\begin{equation}
\les \begin{array}{c} t_s \vspace{4pt} \\
t_p \end{array} \ris = \les \begin{array}{cc} t_{ss}  &
t_{sp} \vspace{4pt} \\ t_{ps} & t_{pp}
\end{array} \ris
 \les \begin{array}{c} a_s \vspace{4pt} \\
a_p \end{array} \ris
\end{equation}
wherein  the reflection coefficients $r_{ss,sp,ps,pp}$ and
transmission coefficients $t_{ss,sp,ps,pp}$ are introduced. The
square magnitude of a reflection   coefficient provides the
corresponding reflectance; i.e.,
\begin{equation}
R_{m n  } = \left| r_{m n} \right|^2,\qquad m, n \in \lec s, p\ric,
\end{equation} while the four transmittances are specified by
\begin{equation}
T_{m n} = \frac{\sqrt{\eps_t} \; \mbox{Re} \, \les \, \cos
\theta_{t} \, \ris}{\sqrt{\eps_{i}} \, \cos \theta_{i}}\,
 \left|
t_{m n} \right|^2,\qquad m , n \in \lec s, p\ric.
\end{equation}

 The absorbances for incident plane waves of the $p$- and
$s$-polarization states   are defined as
\begin{equation}
\left.
\begin{array}{l}
A_p = 1 - \le R_{pp} + R_{sp} +  T_{pp} + T_{sp} \ri  \vspace{6pt}\\
A_s = 1 - \le R_{ss} + R_{ps} +  T_{ss} + T_{ps} \ri
\end{array}
\right\}.
\end{equation}
These absorbances are
 used to identify SPP waves in the modified
 Kretschmann configuration. A sharp high peak in the graph of an
absorbance versus $\theta_{i}$, occurring at $\theta_{i} =
\theta^\sharp_{i}$ say, is a distinctive characteristic of SPP
excitation at the $z=L_m$ interface provided that
\begin{itemize}
\item[ (i)] $\theta^\sharp_{i}$ does not vary when
 the  thickness of the dielectric slab (i.e., $L_\Sigma - L_m$) changes beyond a minimum, and
\item[ (ii)] all four eigenvalues of the matrix $ \les \#Q
(\eps'_d, \kappa', \omega' ) \ris $
 evaluated  at
$\theta^\sharp_{i}$ have non--zero imaginary parts.
\end{itemize}

\section{Numerical studies}\l{ns}

Let us  explore by numerical means the effects of relative motion on
the propagation of SPP waves guided by the planar interface $z =
L_m$. We fix the free-space wavelength in the laboratory frame $\Pi$
at 633~nm. Next, we choose $\eps_i = 6.656$ which is the relative
permittivity of zinc selenide, and for simplicity we take $\eps_t =
\eps_i$.  For the relative permittivity of the dielectric slab we
choose $\eps'_d = 2$, while for the metal film we choose $\eps_m =
-56 + 21i$ which is the relative permittivity of aluminum. The
dielectric slab is taken to have a thickness of 1000 nm, while the
metal film has a thickness of 15 nm; i.e., $L_\Sigma = 1015$ nm and
$L_m = 15$ nm.

 Before considering the effects of relative motion, we
must first establish the baseline for our studies, which is
represented by the scenario wherein there is no relative motion. In
Fig.~\ref{fig2}, the absorbances $A_p$ and $A_s$ are plotted versus
angle of incidence for the case $\beta = 0$. Also plotted on these
graphs are the quantities $R_p = R_{pp} + R_{sp}$ and $R_s = R_{ss}
+ R_{ps}$, calculated when the metal film is absent. A sharp  high
peak in the graph of $A_p$ at $\theta_i = 34.2^\circ$~---~which lies
in the angular regime ${\cal R}_p$ for total reflection in the
absence of the metal film~---~is the signature of SPP excitation by
a $p$-polarized incident plane wave; for the chosen parameters,
${\cal R}_p=\lec\theta_i \vert\theta_i>33.5^\circ\ric$. This
identification is further confirmed by the facts that (i) the
$A_p$-peak remains at $\theta_i = 34.2^\circ$  when the thickness of
the dielectric slab is increased beyond 1000~nm and (ii) the
   eigenvalues of  $ \les \#Q
(\eps'_d, \kappa', \omega' ) \ris $
 evaluated  at
$\theta_i = 34.2^\circ$ all have non--zero imaginary parts. There is
no corresponding sharp high peak in the graph of $A_s$ in
Fig.~\ref{fig2}, in accordance with standard results \c{AZL2,Kret},
although an angular regime ${\cal R}_s$ for total reflection in the
absence of the metal film does exist; note that ${\cal R}_p\equiv
{\cal R}_s$ when $\beta=0$.

\subsection{Moving dielectric slab}

Now we investigate  the scenario schematically depicted in
Fig.~\ref{fig1}, where the dielectric slab moves at constant
velocity parallel to the $x$ axis; i.e., $\hat{\#v} = \hat{\#x}$.
The absorbance $A_p$ is plotted versus $\theta_i$ for the relative
speeds  $\beta \in \lec 0.3, 0.6, 0.8, 0.85, 0.86, 0.9 \ric $ in
Fig.~\ref{fig3}. To confirm whether any possible SPP peaks occur in
the angular regime ${\cal R}_p$, the quantity $R_p = R_{pp} +
R_{sp}$, calculated when the metal film is absent, is also plotted.

The SPP peak in the plots of $A_p$ is found to arise at lower values
of $\theta_i$ when $\beta$ increases from zero. Furthermore,  the
SPP peak in the plot of $A_p$ observed for $\beta \leq 0.85$ is
joined by another prominent peak when  $\beta > 0.85$. For example,
at $\beta = 0.9$, there are prominent $A_p$-peaks at $\theta_{i} =
\theta^{\sharp 1}_{i}= 24.2^\circ$  and at $\theta_{i} =
\theta^{\sharp 2}_{i}= 45.7^\circ$. These values of $\theta^{\sharp
1}_{i}$ and $\theta^{\sharp 2}_{i}$ do not vary when the thickness
of the dielectric slab is increased beyond 1000 nm; and  all four
eigenvalues of the matrix $ \les \#Q (\eps'_d, \kappa', \omega' )
\ris $
 evaluated  at
$\theta^{\sharp 1}_{i}$ and $\theta^{\sharp 2}_{i}$ have non--zero
imaginary parts.

The plots of $R_p$ for $L_m=0$ in Fig.~\ref{fig3} indicate that the
total reflection is not possible at $\beta > 0.85$
  for sufficiently large angles of incidence and therefore the significance
of the two $A_p$-peaks at $\theta^{\sharp 1}_{i}$ and $\theta^{\sharp
2}_{i}$ is unclear. However,
 the $A_p$-peak at
 $\theta^{\sharp 1}_{i}$ corresponds to the SPP peak observed in
the plots of $A_p$ for $\beta \leq 0.85$, as can be confirmed by
tracking this peak while $\beta$ is continuously varied.
 The plots
of $A_s$ (not shown here) corresponding to those of Fig.~\ref{fig3}
for $ \beta \in \le 0, 1 \ri$ do not exhibit SPP peaks.

Next we consider the scenario where the dielectric slab moves at
constant velocity parallel to the $y$ axis; i.e., $\hat{\#v} =
\hat{\#y}$. The absorbance $A_p$, and the quantity $R_p$ calculated
when the metal film is absent, are plotted
 versus $\theta_i$ for the relative speeds  $\beta \in \lec 0.3,
0.6, 0.7, 0.8 \ric $ in Fig.~\ref{fig4}. Unlike for the case of
$\hat{\#v} = \hat{\#x}$ in Fig.~\ref{fig3}, now the value of
$\theta_i$ at which  the $A_p$-peak indicating the excitation of an
SPP wave occurs increases as $\beta$ increases. Furthermore, this
peak   vanishes for $\beta \gtrsim 0.7$.

The corresponding plots of absorbance for an $s$-polarized incident
plane wave, provided in Fig.~\ref{fig5}, exhibit a sharp peak for
mid--range values of $\beta$, at angles of incidence in the angular
regime ${\cal R}_s$. For example,   this $A_s$-peak arises at
$\theta_i = 47.2^\circ$ for $\beta = 0.6$. Further calculations have
revealed that this peak is insensitive to changes in the thickness
of the dielectric slab and all
   eigenvalues of  $ \les \#Q
(\eps'_d, \kappa', \omega' ) \ris $
 evaluated  at
$\theta_i = 47.2^\circ$ have non--zero imaginary parts.

We therefore infer that, in the case of $\hat{\#v} = \hat{\#y}$, the
SPP wave is excited by an incident $p$-polarized plane wave for low
values of $\beta$, but excited by an incident $s$--polarized plane
wave at higher values of $\beta$. As $\beta$ approaches unity,  no
absorbance peak indicating the excitation of an SPP wave is
observed, for incident plane waves of either linear polarization
state.

Clearly, from Figs.~\ref{fig3}--\ref{fig5}, the direction of motion
has a major bearing on the excitation of SPP waves. This sensitivity
to direction of motion  reflects the fact that $\kappa \ne\kappa'$
when motion is parallel to the plane of incidence but $\kappa =
\kappa'$ when motion is perpendicular to the plane of incidence, as
may be inferred from \r{k_LT}.

In order to further illuminate this
issue, let us investigate the case where the direction of motion is
neither parallel nor perpendicular to the plane of incidence; i.e.,
$\hat{\#v} = \hat{\#x}\cos\psi+\hat{\#y}\sin\psi$ where, for
example, we choose the angle $\psi = 45^\circ$. In  Fig.~\ref{fig6},
 the absorbance
$A_p$, and the quantity $R_p$ calculated when the metal film is
absent, are plotted
 versus $\theta_i$ for the relative speeds  $\beta \in \lec 0.3,
0.6, 0.8, 0.88, 0.9, 0.95 \ric $. The plots in Fig.~\ref{fig6}
display some features of the plots presented in both
Figs.~\ref{fig3} and \ref{fig4}. As is the case when motion is
parallel to the plane of incidence, the solitary $A_p$-peak observed
at low and moderate values of $\beta$ is joined by a second peak at
higher values of $\beta$. A second $A_p$-peak begins to emerge in
Fig.~\ref{fig6}  at $\beta \approx 0.88$; it is fully developed at
$\beta = 0.9$; and as $\beta$ increases to approximately 0.95, the
second $A_p$-peak coalesces with the first $A_p$-peak. For example,
at $\beta = 0.9$, there are $A_p$-peaks at $\theta_i =
\theta^{\sharp 1}_{i} = 32.3^\circ$ and at $\theta_i =
\theta^{\sharp 2}_{i} = 64.2^\circ$. The peak at $ \theta^{\sharp
1}_{i}$ corresponds to the solitary SPP peak which is observed at
low and moderate values of $\beta$. As is the case for the two
high--$\beta$ $A_p$-peaks in Fig.~\ref{fig3}, the values of
$\theta^{\sharp 1}_{i}$ and $\theta^{\sharp 2}_{i}$ do not vary when
the thickness of the dielectric slab is increased beyond 1000 nm;
and  all four eigenvalues of the matrix $ \les \#Q (\eps'_d,
\kappa', \omega' ) \ris $
 evaluated  at
$\theta^{\sharp 1}_{i}$ and $\theta^{\sharp 2}_{i}$ have non--zero
imaginary parts. We note that the $A_p$-peak at  $\theta^{\sharp 1}_{i}$ is
considerably  sharper than the $A_p$-peak at  $\theta^{\sharp 2}_{i}$.

The manifestation of an $A_s$-peak at moderate values of $\beta$, as
may be observed in Fig.~\ref{fig5} when the motion is directed
perpendicular to the plane of incidence, also occurs when motion is
directed along $\hat{\#v} = \le \hat{\#x} +\hat{\#y}\ri / \sqrt{2}$,
as is demonstrated by the plots presented in  Fig.~\ref{fig7}.
Furthermore, the $A_s$ plots in Fig.~\ref{fig7}  exhibit  two $A_s$-peaks
at high values of $\beta$, which coalesce as $\beta $ approaches
unity, in a similar manner to the two $A_p$-peaks observed in
Fig.~\ref{fig6}.

\subsection{Moving metal film}
Next, we turn to a rather different situation. The dielectric slab
is now held fixed relative to the half--spaces $z < 0$ and $z >
L_\Sigma$, whereas the metal film  moves with velocity $\#v = \beta
\co \hat{\#v}$ in the $xy$ plane. The corresponding formulation is
isomorphic to that presented in Sec.~\ref{analysis}, but with the
treatments for the metal film and the dielectric slab interchanged.
For the case $\hat{\#v} = \hat{\#x}$, plots of $A_p$ versus
$\theta_i$ (not shown here) are largely insensitive to $\beta$. That
is, the $A_p$-peak indicating the excitation of an SPP wave occurs
at approximately the same value of $\theta_i$ and has approximately
the same amplitude, regardless of $\beta$. Furthermore, there are no
notable peaks in the corresponding plots of $A_s$.

However, plots of $A_p$ in the case of $\hat{\#v} = \hat{\#y}$ are
much more sensitive to $\beta$. As we can  see in Fig.~\ref{fig8},
the $A_p$-peak indicating the excitation of an SPP wave occurs at
slightly higher values of $\theta_i$ as $\beta$ increases. Also, the
amplitude of this peak gradually diminishes as $\beta$ increases, to
such an extent that the peak is barely discernible at $\beta = 0.9$.
In the corresponding plots for a $s$--polarized incident plane wave
(not shown here), there is a mere hint of an $A_s$-peak indicating
the excitation of an SPP wave  at mid--range values of $\beta$, but
nothing as obvious as appears in the $s$-polarization scenario
represented in Fig.~\ref{fig5}.

\section{Discussion}\l{cr}

Uniform motion can have a major effect on the excitation of SPP
waves guided by the interface of a metal film and a dielectric slab,
both isotropic and homogeneous in their respective co-moving
inertial reference frames. In the case of a uniformly moving
dielectric slab, the angle of incidence for SPP-wave excitation is
highly sensitive to the relative speed   $\beta$ and the direction
of motion. For the specific example considered in Sec.~\ref{ns},
\begin{itemize}
\item[(a)] when the direction of motion is
 parallel to the
plane of incidence, the SPP wave is excited by $p$-polarized (but
not $s$-polarized) incident plane waves
for low and moderate values
of $\beta$; and
\item[(b)] when the direction of motion is
 perpendicular to the
plane of incidence, the SPP wave is excited by $p$-polarized
incident plane waves for low values of $\beta$, but $s$-polarized
incident plane waves at moderate values of $\beta$,  while at higher
values of $\beta$ the SPP wave is not excited.
\end{itemize}

 Some insight into this sensitivity to relative motion can be gained by
considering the Minkowski constitutive relations for the uniformly
moving dielectric slab. That is, the electromagnetic response of the
uniformly moving dielectric slab may be represented in the
laboratory frame by the bianisotropic constitutive relations
\cite[Chap.~8]{Chen}
\begin{equation}
\left. \begin{array}{l} \displaystyle{ \#D (\#r, \omega) = \epso
\eps'_d \, \=\alpha \cdot \#E (\#r, \omega) + \frac{1}{\co} \le \#m
\times \=I \,\ri \cdot \#H(\#r,
\omega)} \vspace{6pt}\\
\displaystyle{\#B (\#r, \omega) = - \frac{1}{\co} \le \#m \times \=I
\,\ri \cdot \#E(\#r, \omega)} +  \muo \, \=\alpha \cdot \#H (\#r,
\omega) \end{array} \right\},
\end{equation}
where
\begin{equation}
\=\alpha = \alpha \, \=I + \le 1- \alpha \ri \, \hat{\#v}\,
\hat{\#v}, \qquad \alpha = \frac{1 - \beta^2}{1 - \eps'_d \beta^2}
\end{equation}
and
\begin{equation}
\#m = m \, \hat{\#v}, \qquad m = \frac{\beta \le \eps'_d - 1 \ri}{1
- \eps'_d \beta^2}.
\end{equation}
Thus,  when the direction of motion is parallel to the plane of
incidence, for example,  the Minkowski constitutive dyadics take the
form
\begin{equation}
\left.
\begin{array}{l}
 \=\alpha  = \hat{\#x} \, \hat{\#x} + \alpha \le \hat{\#y} \, \hat{\#y} + \hat{\#z} \, \hat{\#z}
 \ri \vspace{4pt} \\
 \#m \times \=I = m \le \hat{\#z} \, \hat{\#y} - \hat{\#y} \,
 \hat{\#z} \ri
 \end{array}
 \right\}.
\end{equation}
The Minkowski constitutive scalars $\alpha$ and $m$ for this case
are plotted versus $\beta$ in Fig.~\ref{fig9}. Clearly, the
constitutive scalars are highly sensitive to $\beta$. Indeed, both
$\alpha$ and $m$ become unbounded as $\beta \to 1/ \sqrt{2} =
0.707$. This singularity may be responsible for  the break down of
the total reflection regime and the anomalous second peak in the
plots of $A_p$ and $A_s$ reported in Sec.~\ref{ns} at high values of
$\beta$.

 In the case of a
uniformly moving metal film, the
 sensitivity  to $\beta$ is less obvious. Since the metal considered in Sec.~\ref{ns} is
 dissipative~---~and quite highly, too~---~the Minkowski
 constitutive relations cannot be used here to gain an insight into the
 sensitivity to $\beta$ \c{ML2009}.

 To conclude, our study further extends our understanding of
 electrodynamic processes at planar interfaces, especially relating to SPP waves.

\vspace{10mm}

\noindent {\bf Acknowledgments:} TGM is supported by a  Royal
Academy of Engineering/Leverhulme Trust Senior Research Fellowship.
AL thanks
 the Binder Endowment at Penn State for partial financial
support of his research activities.

\newpage

\begin{figure}[!ht]
\centering
\includegraphics[width=3.0in]{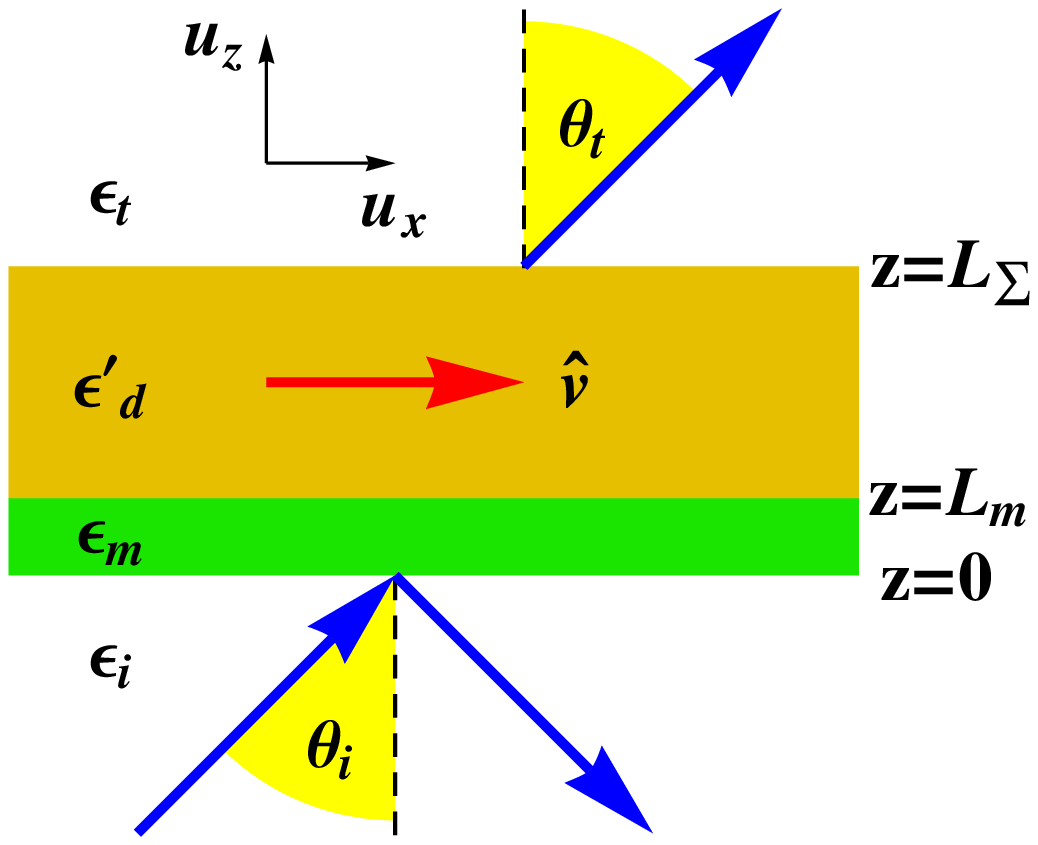}
 \caption{\l{fig1}
 A schematic depiction of the modified Kretschmann configuration under investigation. A
 plane wave is incident upon a metal film occupying $0 < z < L_m$. The region
 $L_m < z < L_\Sigma$ is filled by a dielectric slab
 which moves at constant velocity in the direction of $\hat{\#v}$.
  }
\end{figure}

\vspace{20mm}

\begin{figure}[!ht]
\centering
\includegraphics[width=3.0in]{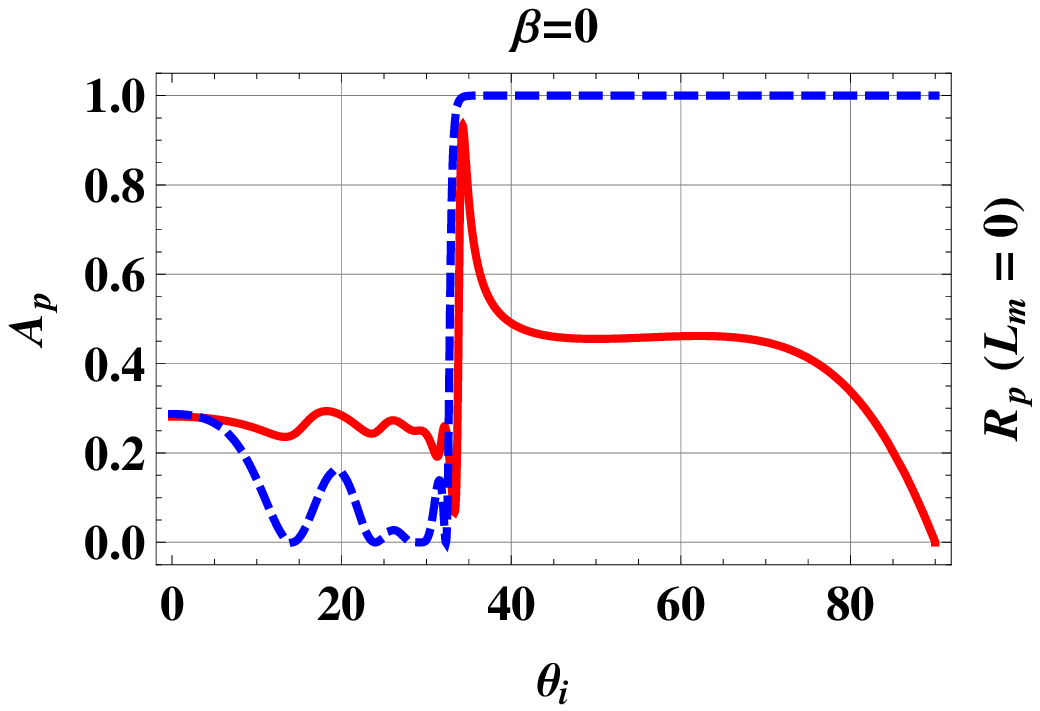}
\includegraphics[width=3.0in]{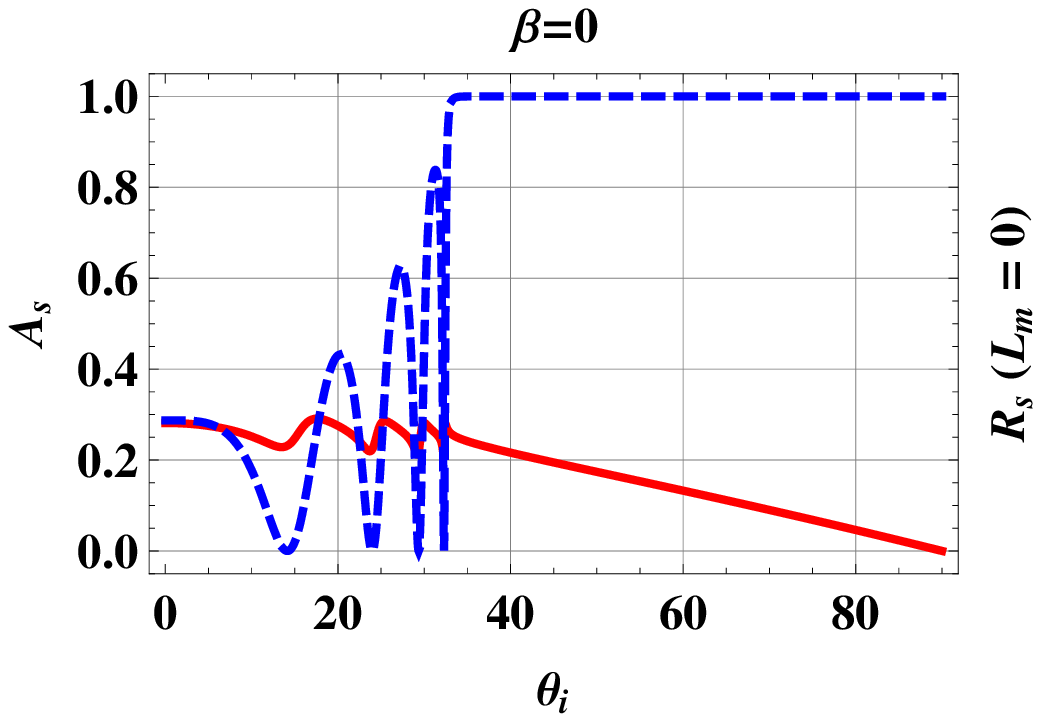}
 \caption{\l{fig2}
 Left: Absorbance $A_p = 1 - (R_{pp} + R_{sp} + T_{pp} + T_{sp})$ (red,
 solid curve) plotted versus
 angle of incidence  $\theta_i$ (in degree) for the case $\beta = 0$;  also plotted is the quantity $R_p =
 R_{pp} + R_{sp}$ (blue, dashed curve), calculated when $L_m =0$.
Right: Absorbance $A_s = 1 - (R_{ss} + R_{ps} + T_{ss} + T_{ps})$
(red,
 solid curve) plotted versus
 angle of incidence $\theta_i$ (in degree) for the case $\beta = 0$;  also plotted is the quantity $R_s =
 R_{ss} + R_{ps}$ (blue, dashed curve), calculated when $L_m =0$.}
\end{figure}

\newpage

\begin{figure}[!ht]
\centering
\includegraphics[width=3.0in]{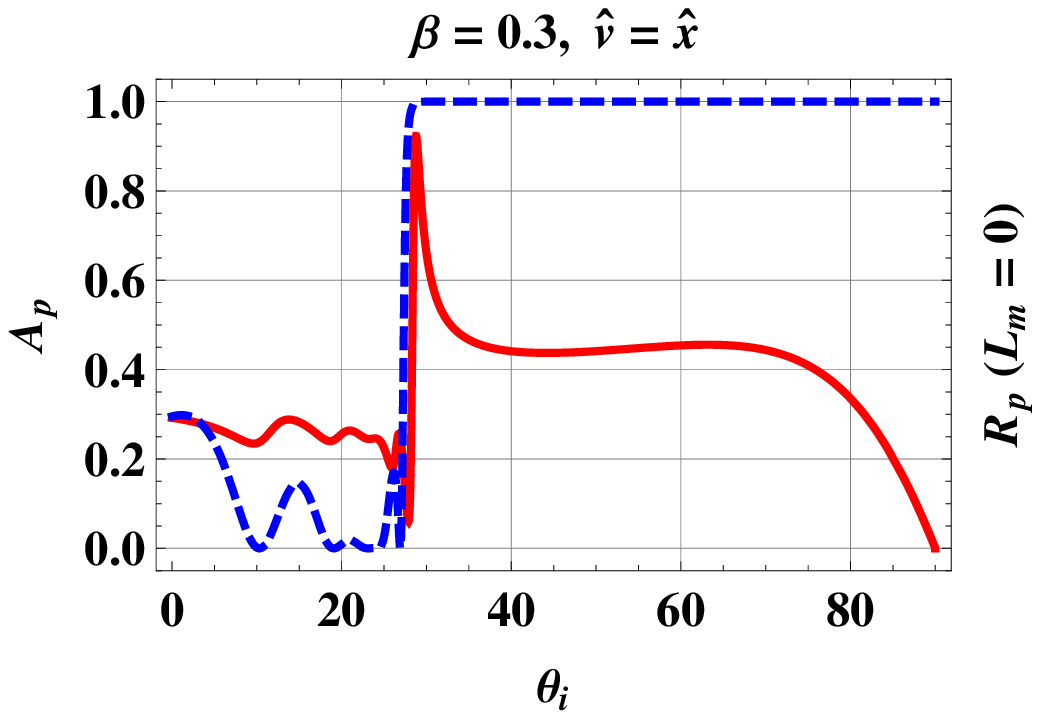}
\hfill
\includegraphics[width=3.0in]{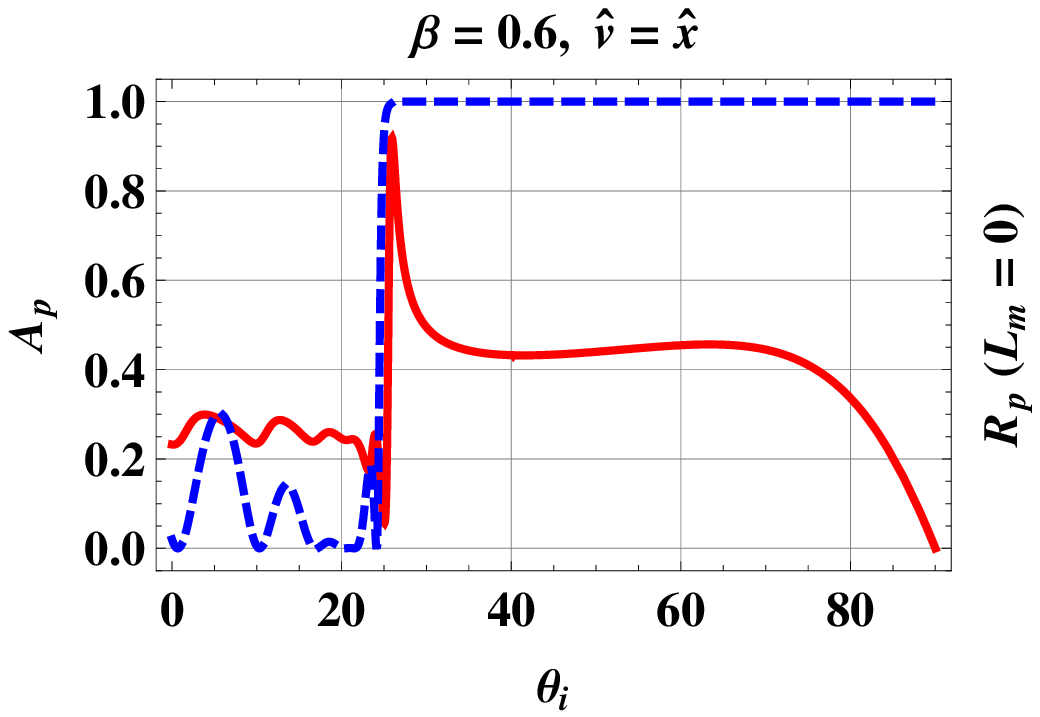} \vspace{10mm} \\
\includegraphics[width=3.0in]{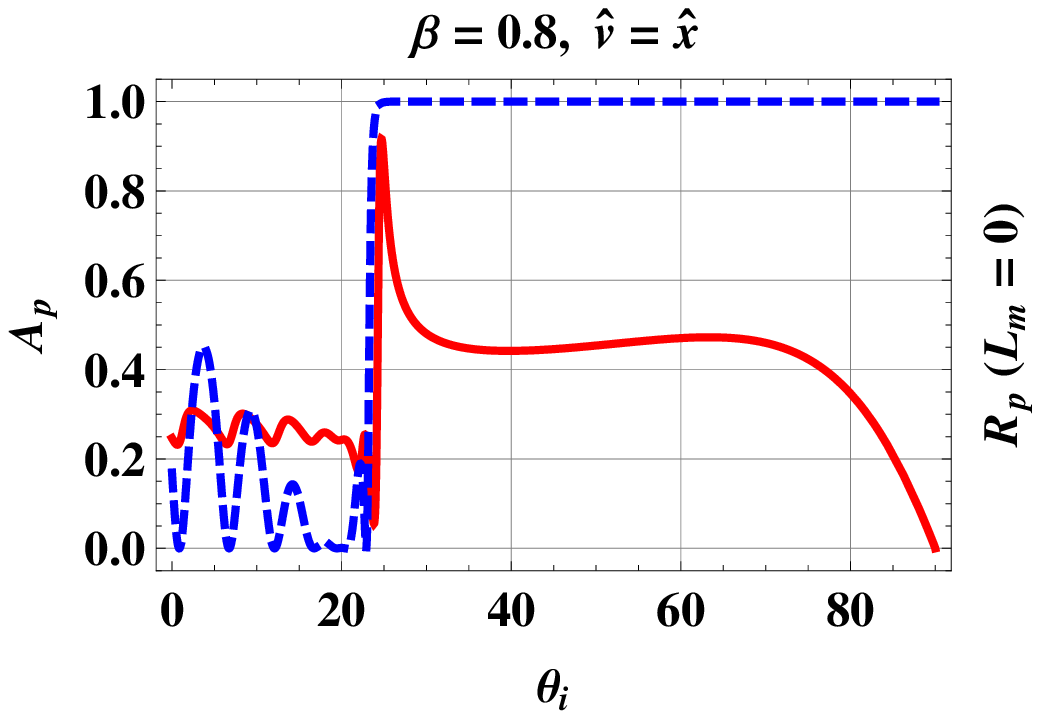}\hfill
\includegraphics[width=3.0in]{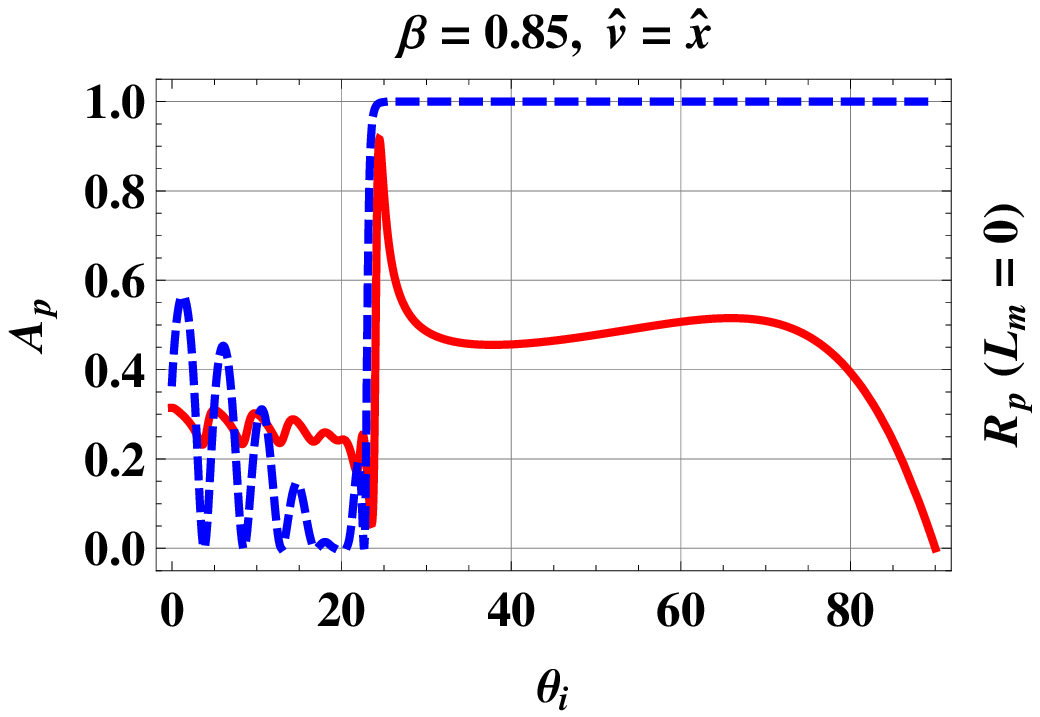}  \vspace{10mm} \\
\includegraphics[width=3.0in]{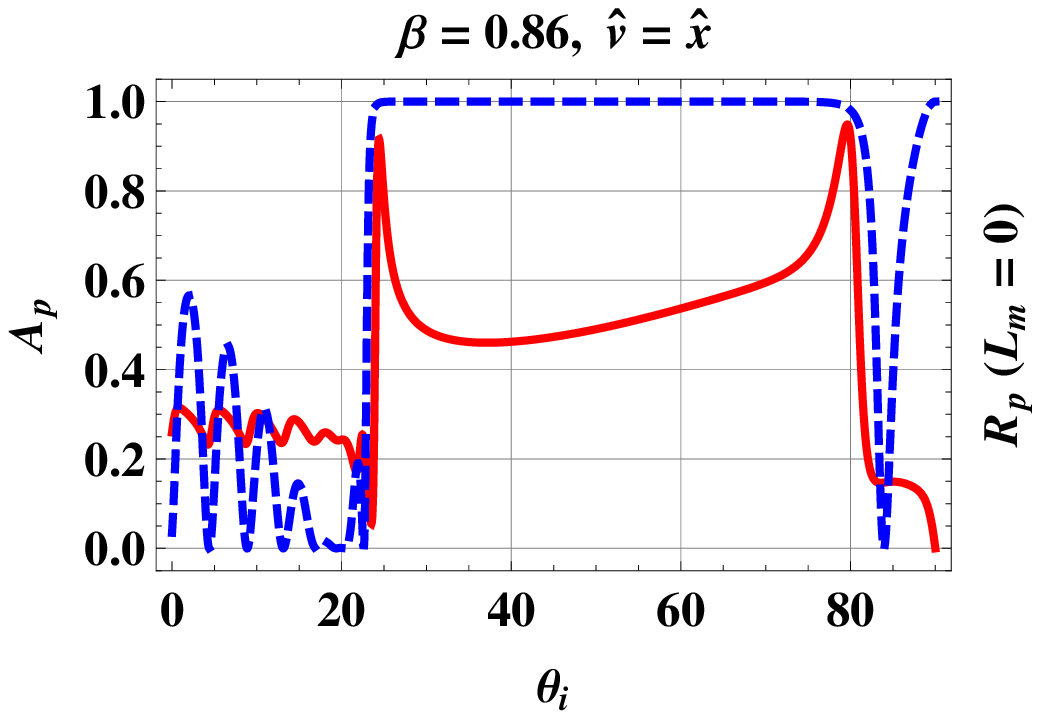}\hfill
\includegraphics[width=3.0in]{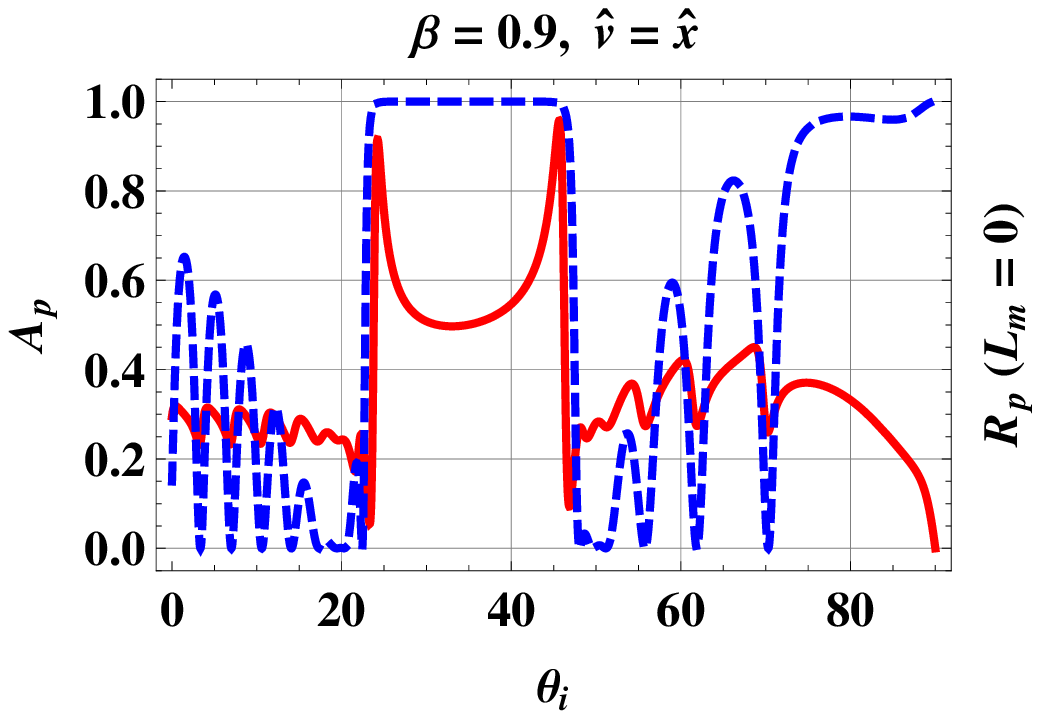}
 \caption{\l{fig3}
 Absorbance $A_p = 1 - (R_{pp} + R_{sp} + T_{pp} + T_{sp})$ (red,
 solid curve) plotted
 versus  angle of incidence $\theta_i$ (in degree) for the scenario
where the dielectric slab moves in the direction
 $\hat{\#v} = \hat{\#x}$ at relative speeds $\beta \in \lec 0.3, 0.6, 0.8, 0.85, 0.86, 0.9 \ric$. Also plotted is the quantity $R_p =
 R_{pp} + R_{sp}$ (blue, dashed curve), calculated when $L_m =0$. }
\end{figure}

\newpage

\begin{figure}[!ht]
\centering
\includegraphics[width=3.0in]{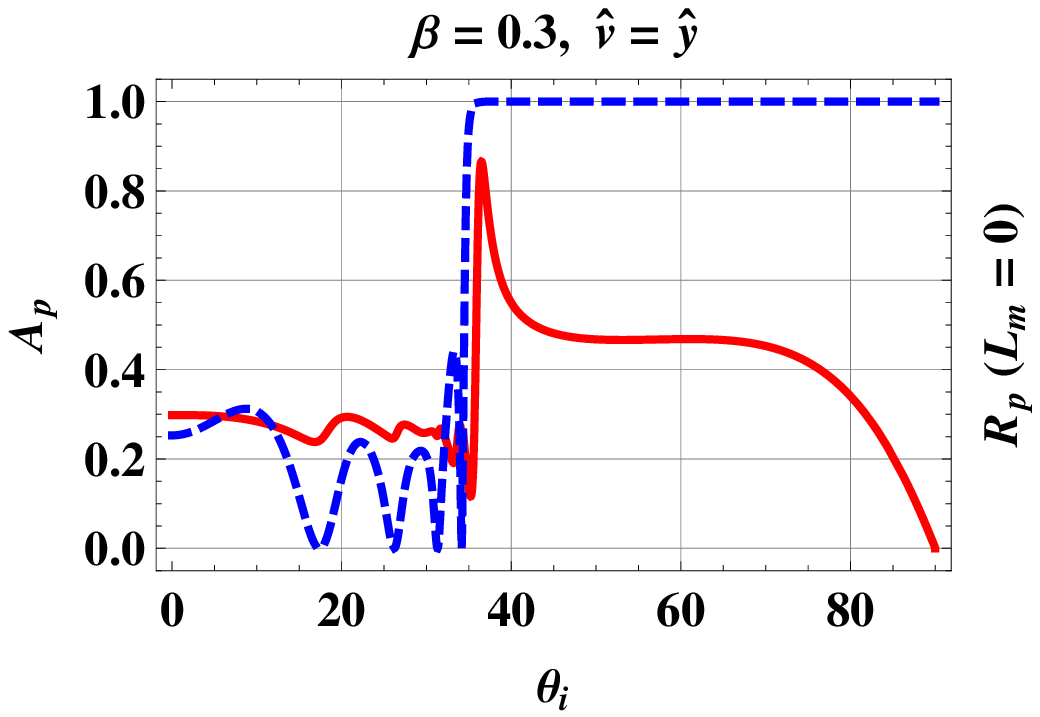} \hfill
\includegraphics[width=3.0in]{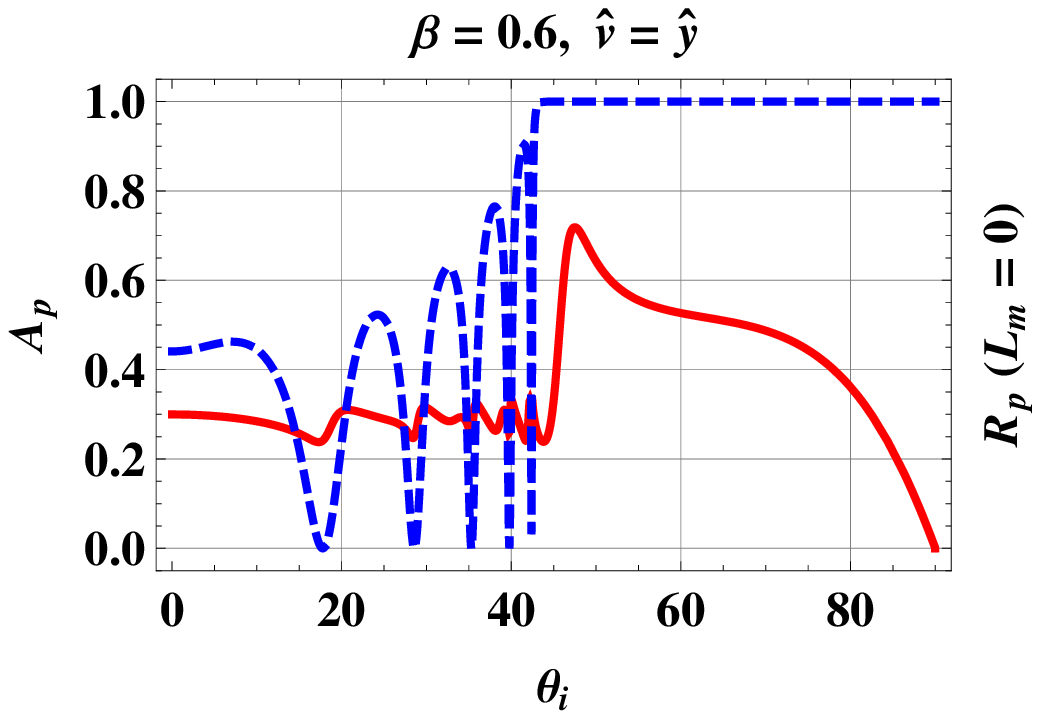} \vspace{10mm} \\
\includegraphics[width=3.0in]{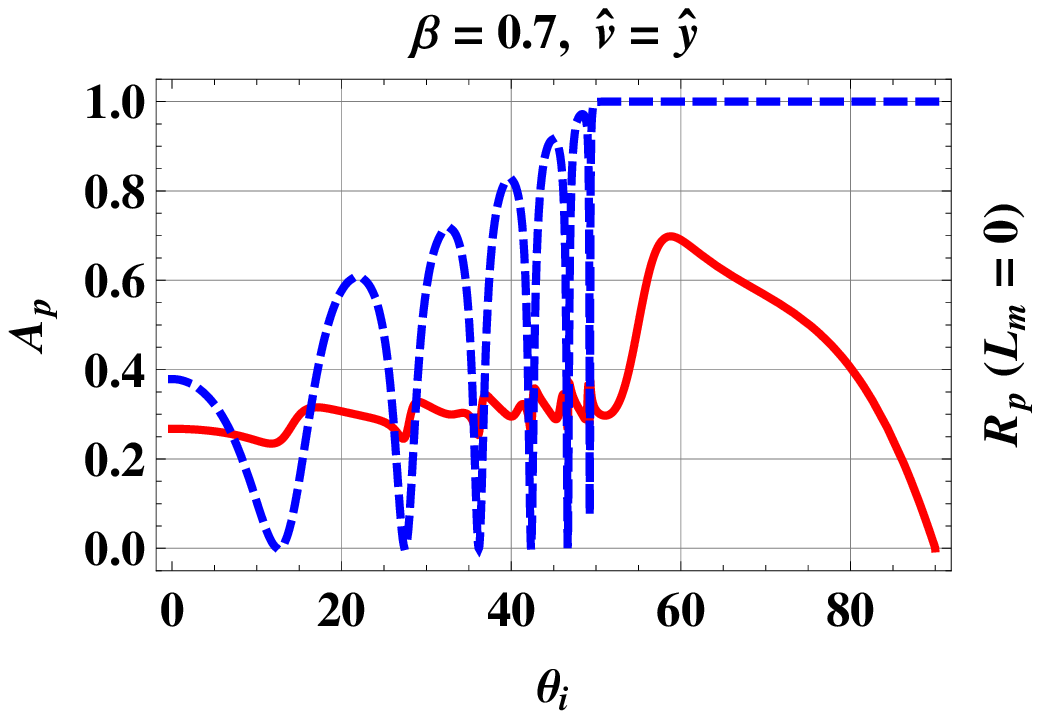} \hfill
\includegraphics[width=3.0in]{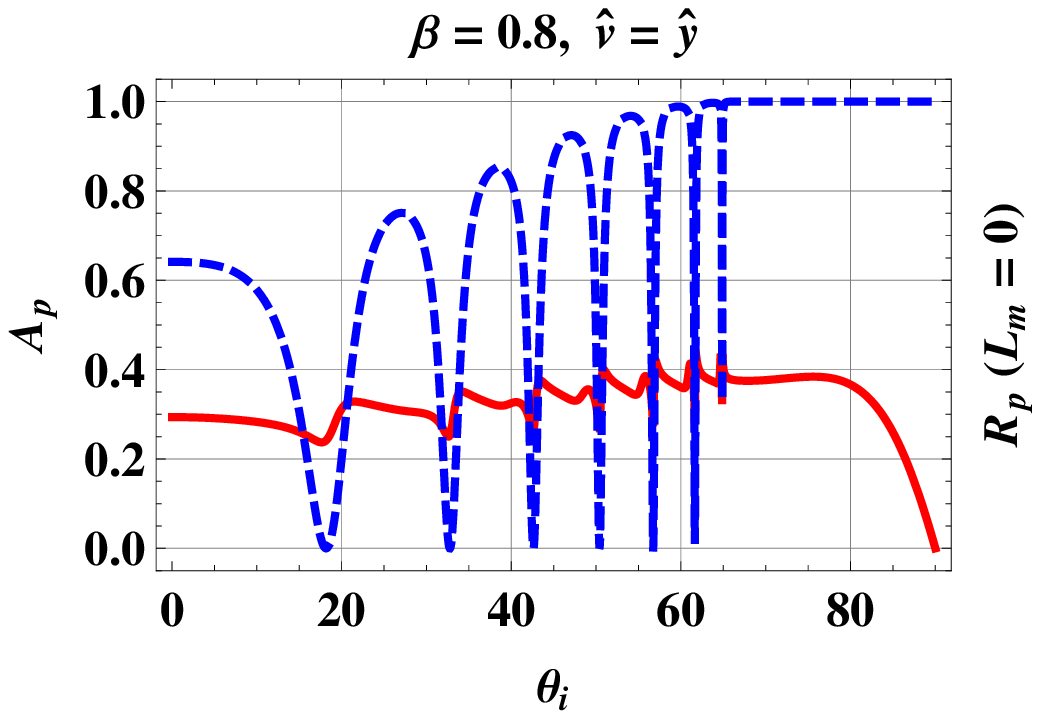}
 \caption{\l{fig4}
 As Fig.~\ref{fig3} except that  the dielectric slab moves in the direction
 $\hat{\#v} = \hat{\#y}$ and $\beta \in \lec 0.3, 0.6, 0.7, 0.8 \ric$. }
\end{figure}

\newpage

\begin{figure}[!ht]
\centering
\includegraphics[width=3.0in]{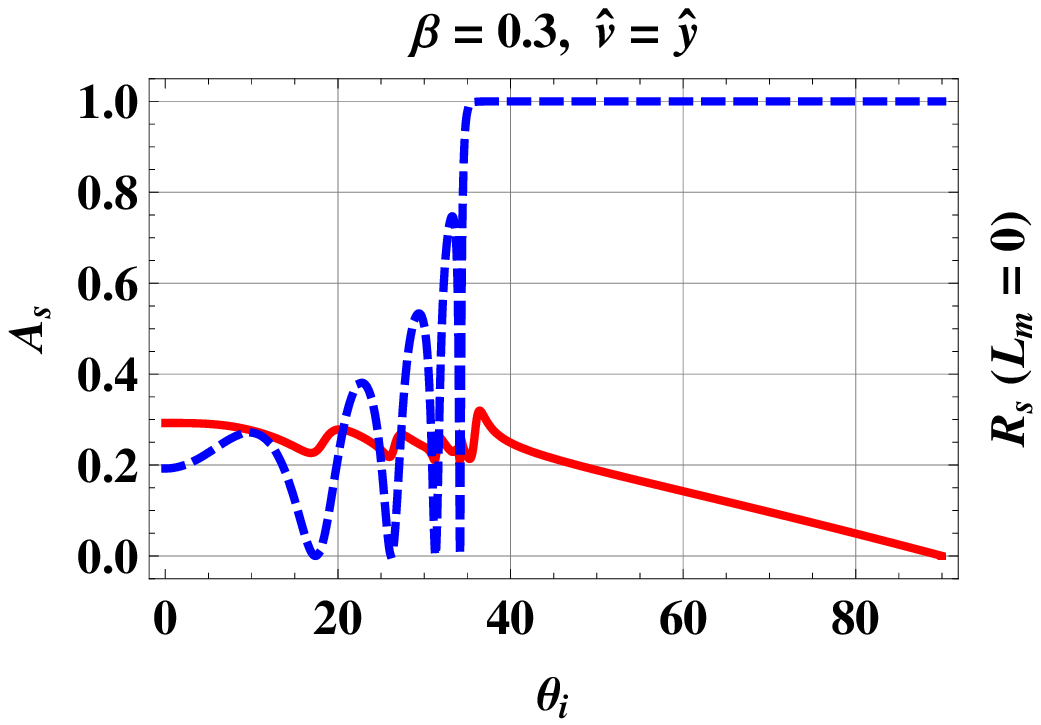} \hfill
\includegraphics[width=3.0in]{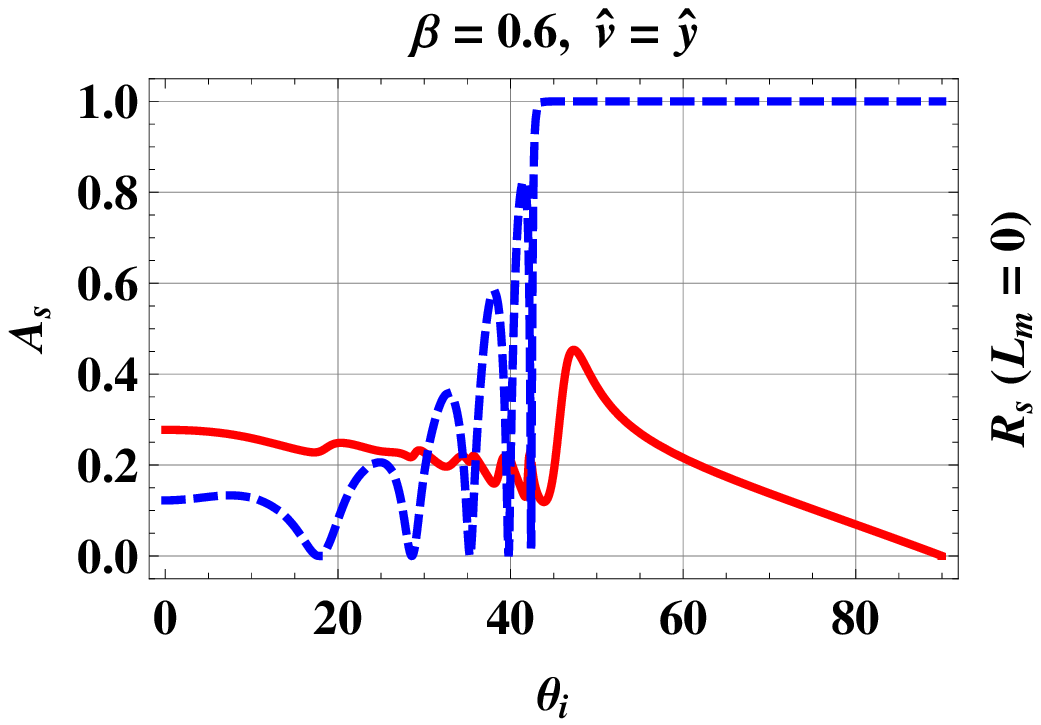} \vspace{10mm} \\
\includegraphics[width=3.0in]{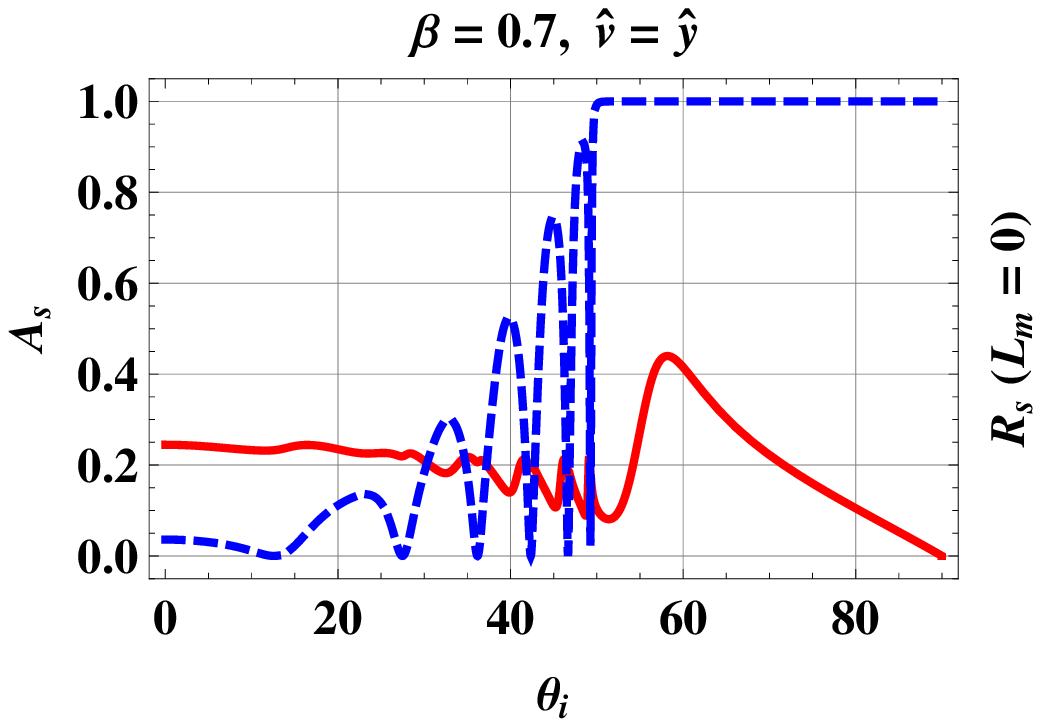} \hfill
\includegraphics[width=3.0in]{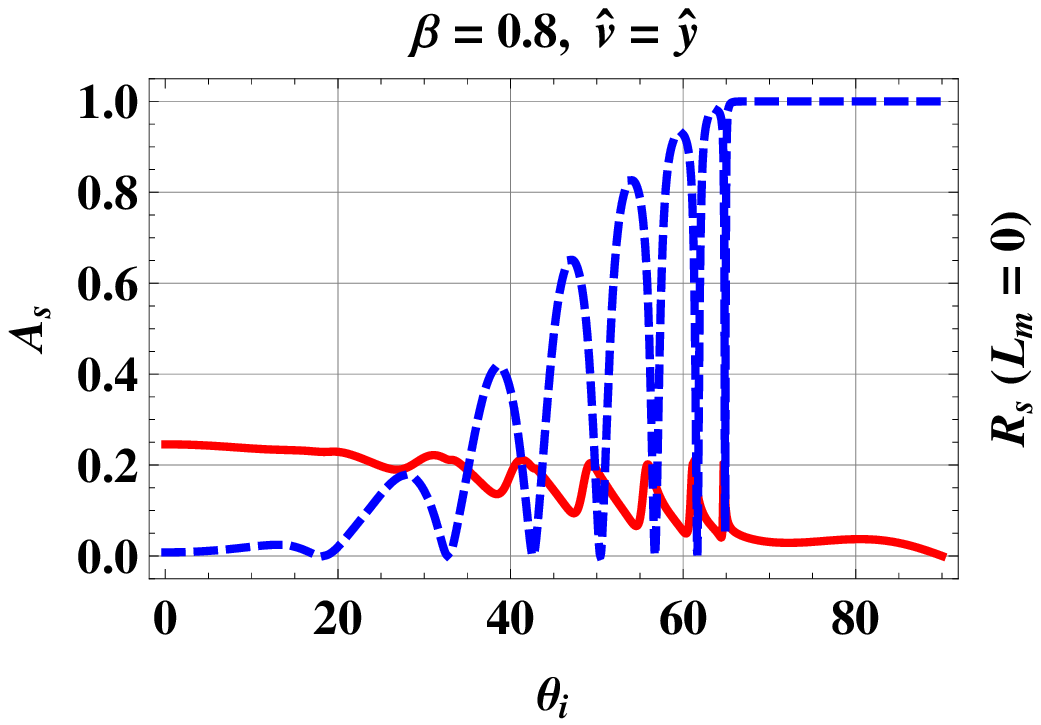}
 \caption{\l{fig5}
 As Fig.~\ref{fig4} except that quantities plotted are $A_s = 1 - (R_{ss} + R_{ps} + T_{ss} + T_{ps})$
(red,
 solid curve)  versus
 angle of incidence $\theta_i$ (in degree), and the quantity $R_s =
 R_{ss} + R_{ps}$ (blue, dashed curve), calculated when $L_m =0$. }
\end{figure}

\newpage

\begin{figure}[!ht]
\centering
\includegraphics[width=3.0in]{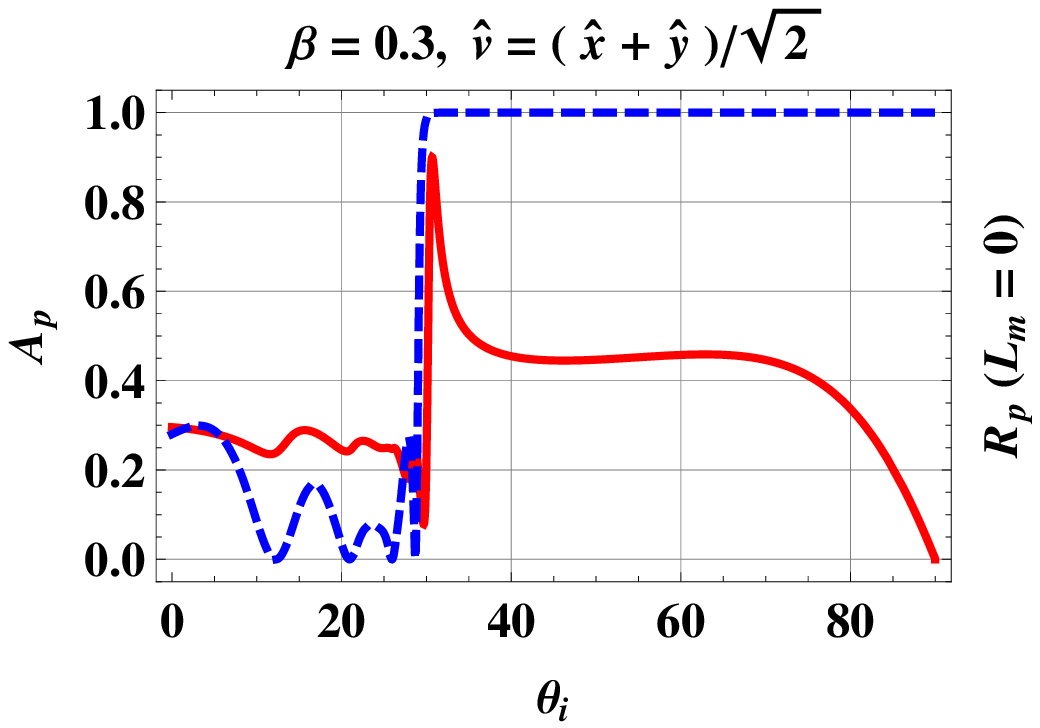} \hfill
\includegraphics[width=3.0in]{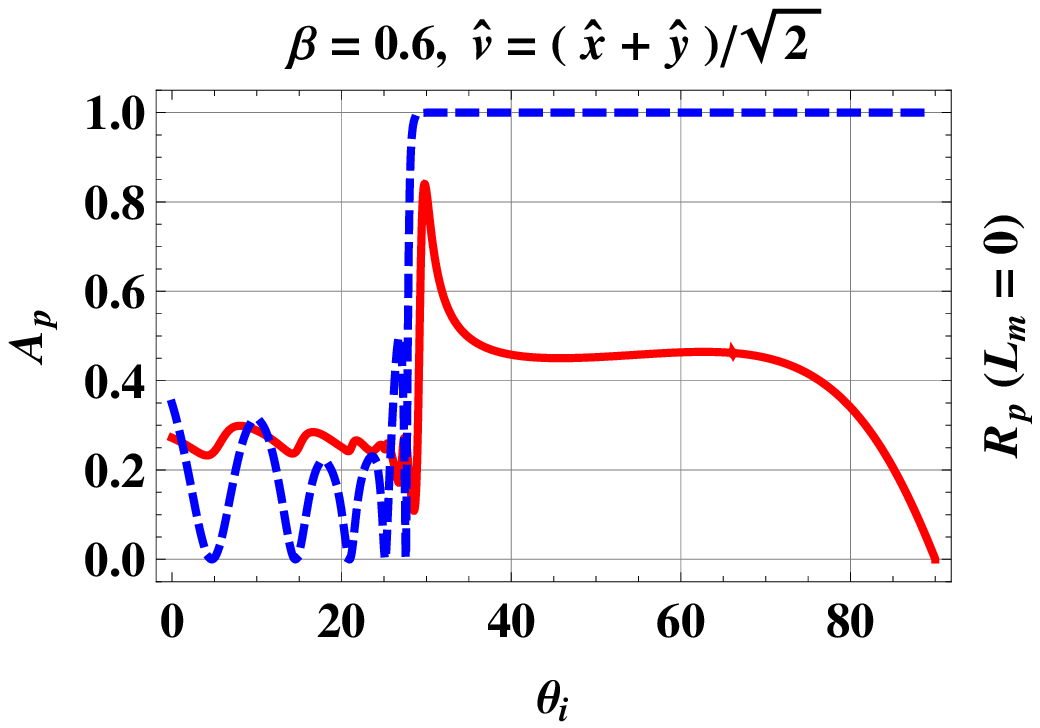}\vspace{10mm} \\
\includegraphics[width=3.0in]{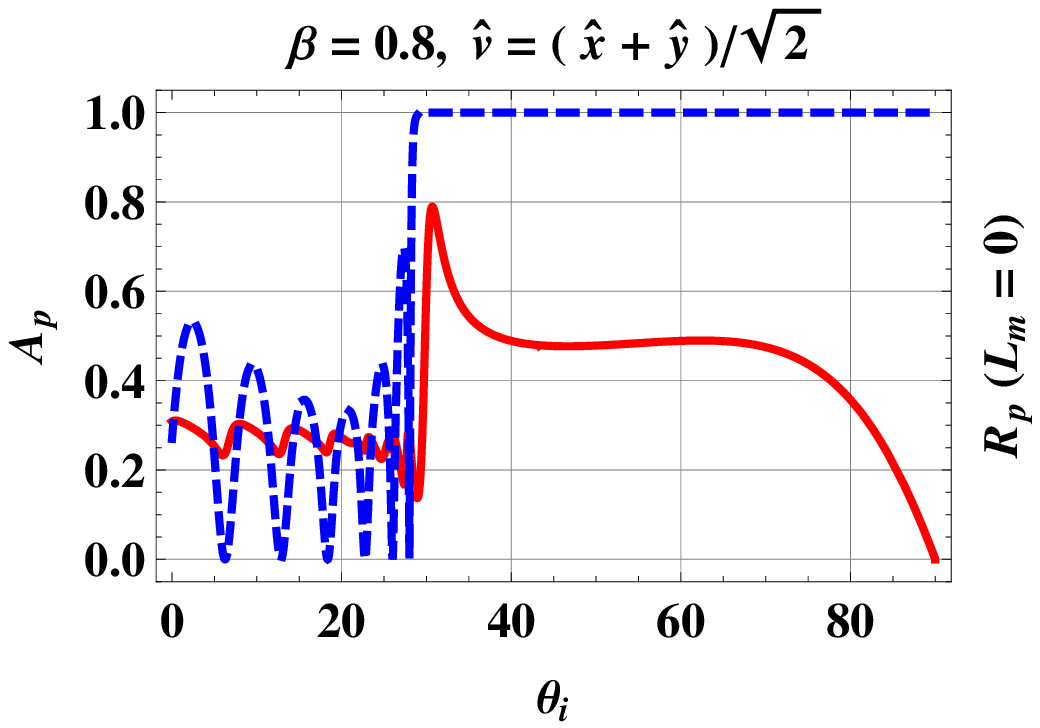} \hfill
\includegraphics[width=3.0in]{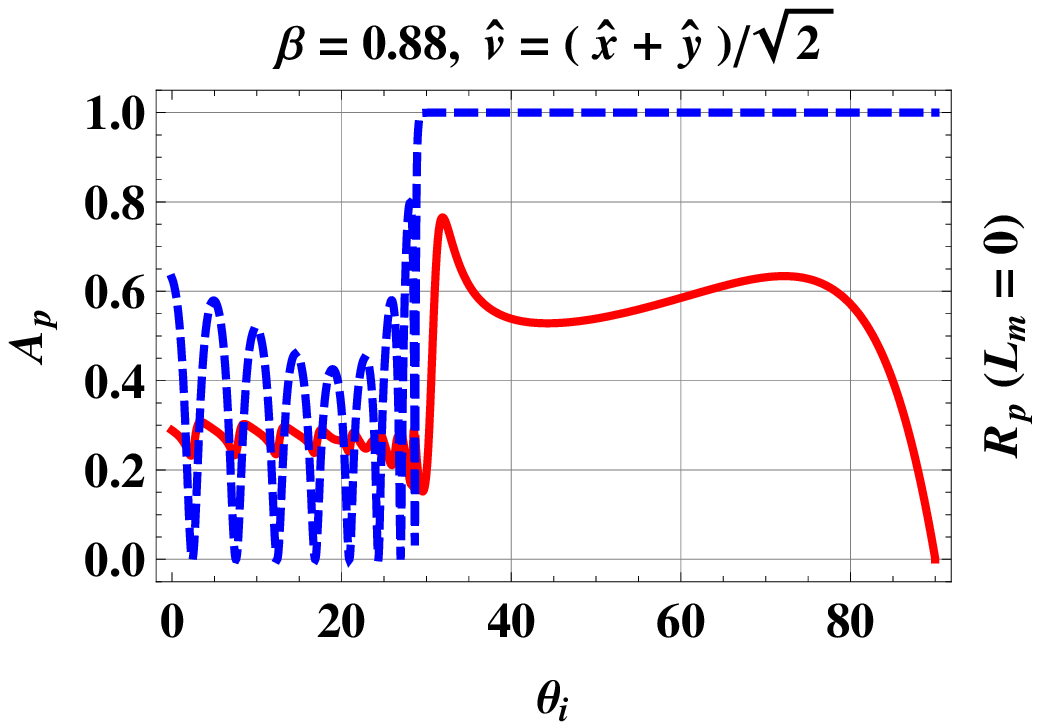} \vspace{10mm} \\
\includegraphics[width=3.0in]{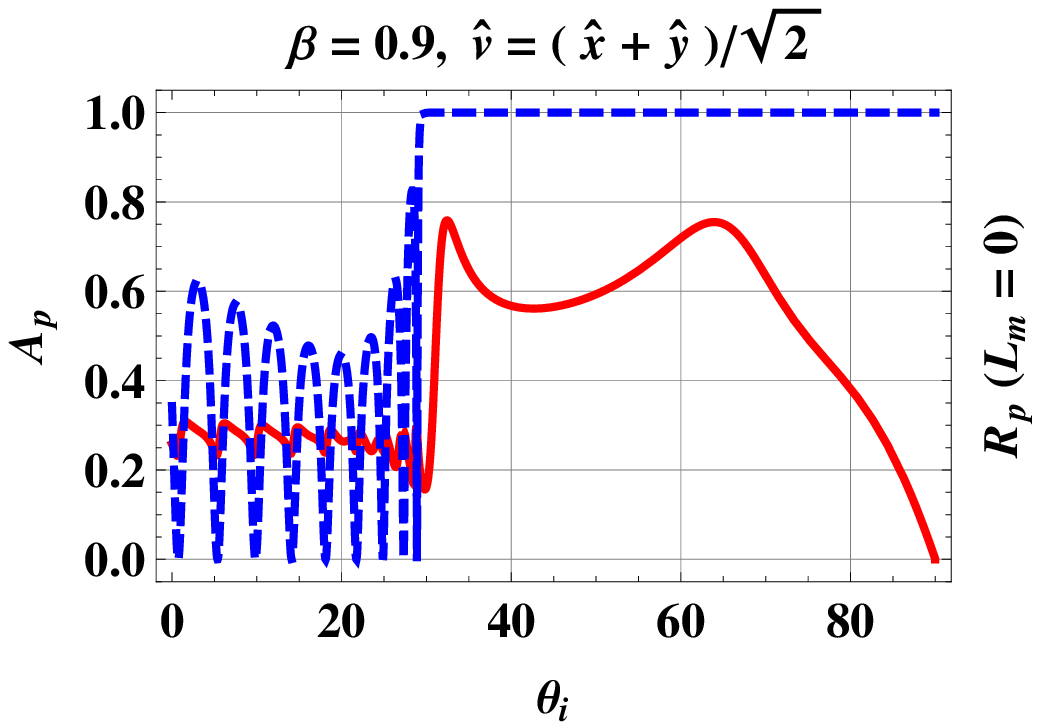} \hfill
\includegraphics[width=3.0in]{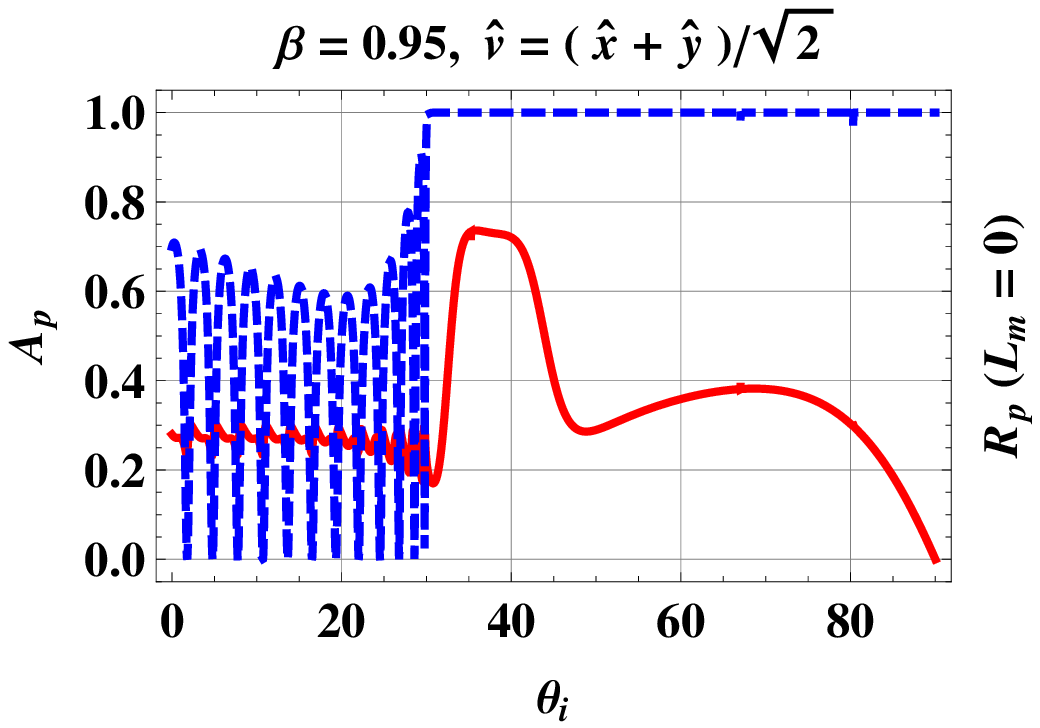}
 \caption{\l{fig6}
 Absorbance $A_p = 1 - (R_{pp} + R_{sp} + T_{pp} + T_{sp})$ (red,
 solid curve) plotted
 versus  angle of incidence $\theta_i$ (in degree) for the scenario
where the dielectric slab  moves in the direction
 $\hat{\#v} = \le \hat{\#x} +  \hat{\#y} \ri / \sqrt{2}$
  at relative speeds $\beta \in \lec 0.3, 0.6, 0.8, 0.88,  0.9, 0.95 \ric$. Also plotted is the quantity $R_p =
 R_{pp} + R_{sp}$ (blue, dashed curve), calculated when $L_m =0$.
 }
\end{figure}

\newpage

\begin{figure}[!ht]
\centering
\includegraphics[width=3.0in]{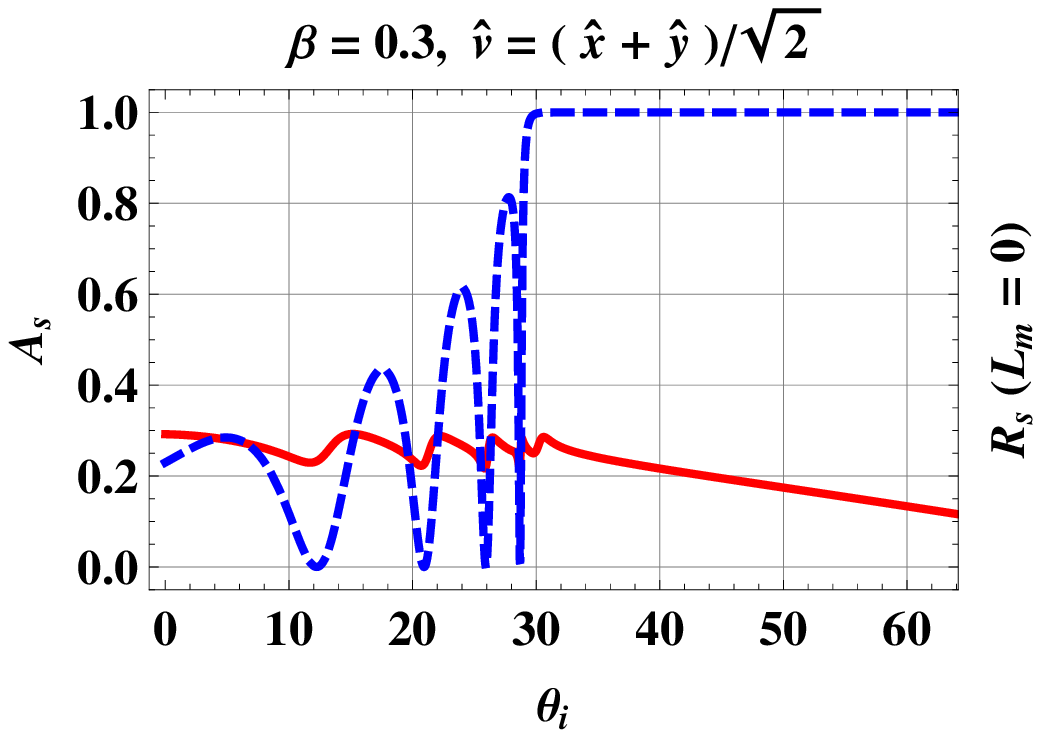} \hfill
\includegraphics[width=3.0in]{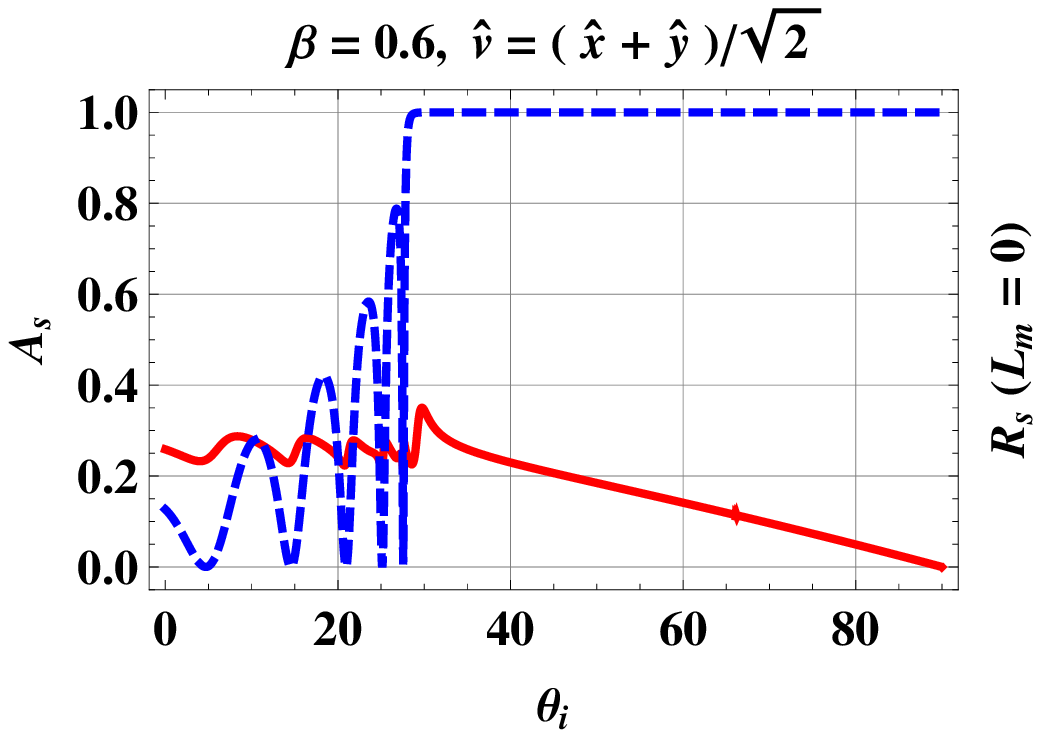}\vspace{10mm} \\
\includegraphics[width=3.0in]{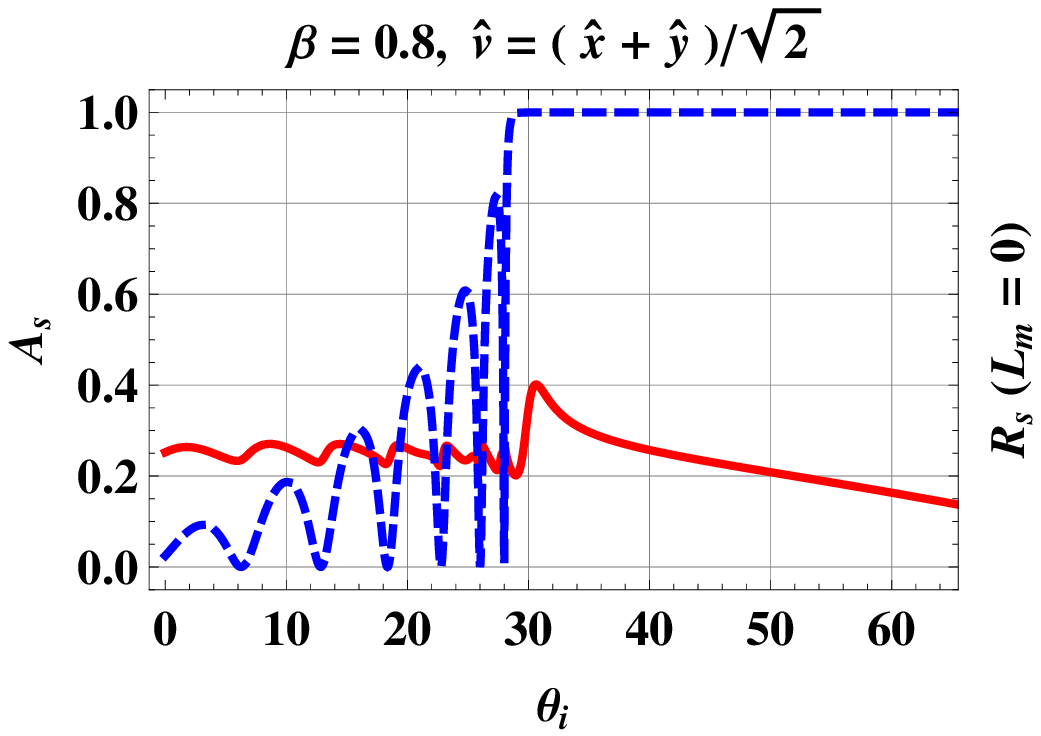} \hfill
\includegraphics[width=3.0in]{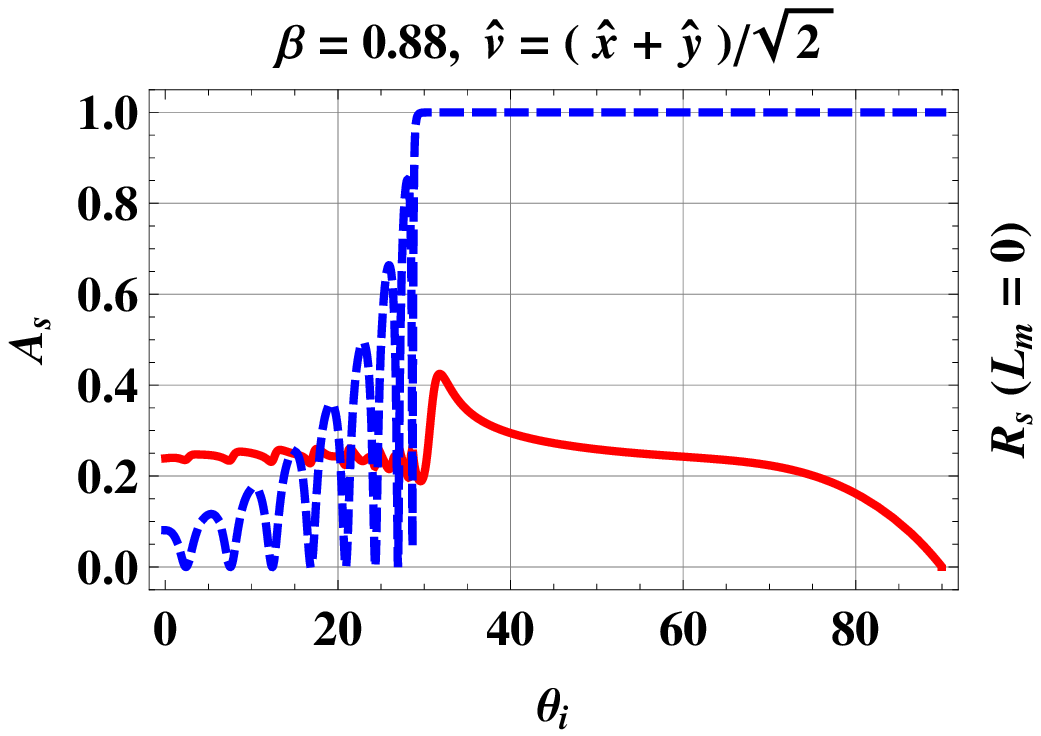} \vspace{10mm}\\
\includegraphics[width=3.0in]{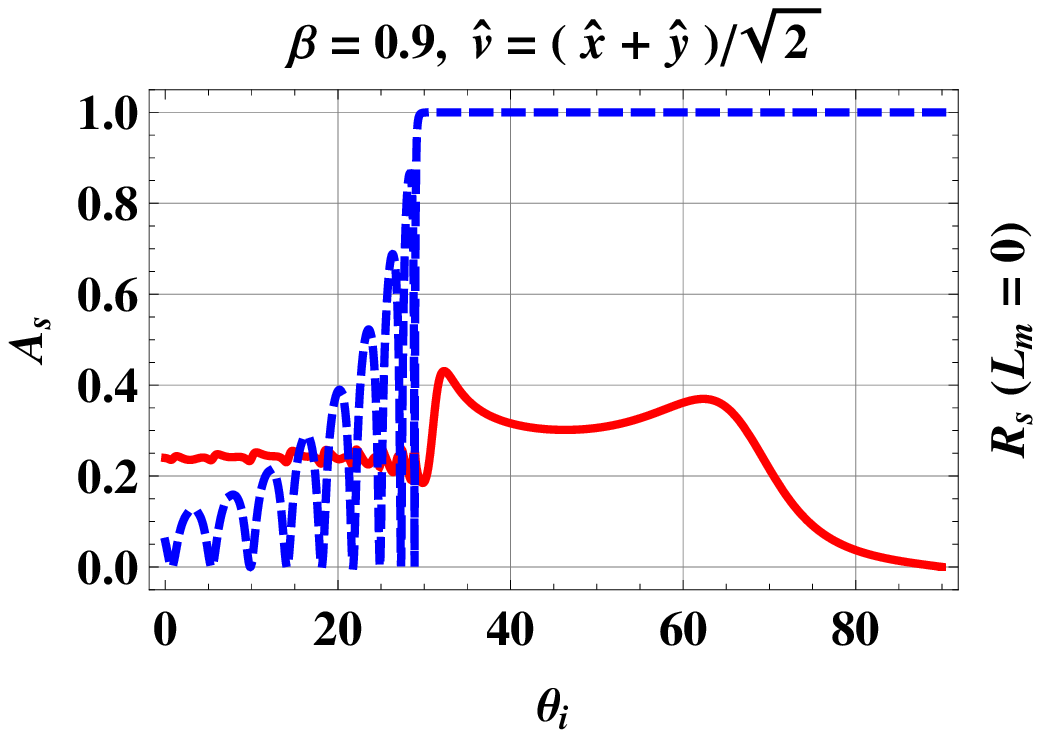} \hfill
\includegraphics[width=3.0in]{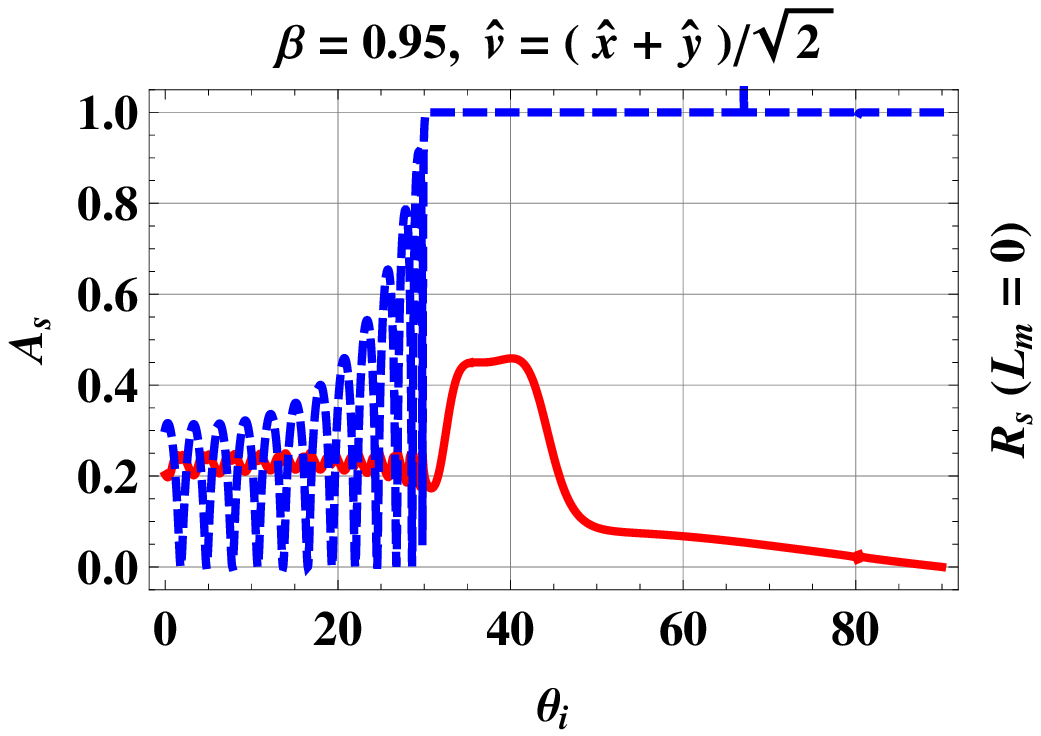}
 \caption{\l{fig7}
 As Fig.~\ref{fig6} except that quantities plotted are $A_s = 1 - (R_{ss} + R_{ps} + T_{ss} + T_{ps})$
(red,
 solid curve)  versus
 angle of incidence $\theta_i$ (in degree), and the quantity $R_s =
 R_{ss} + R_{ps}$ (blue, dashed curve), calculated when $L_m =0$.
 }
\end{figure}

\newpage

\begin{figure}[!ht]
\centering
\includegraphics[width=3.0in]{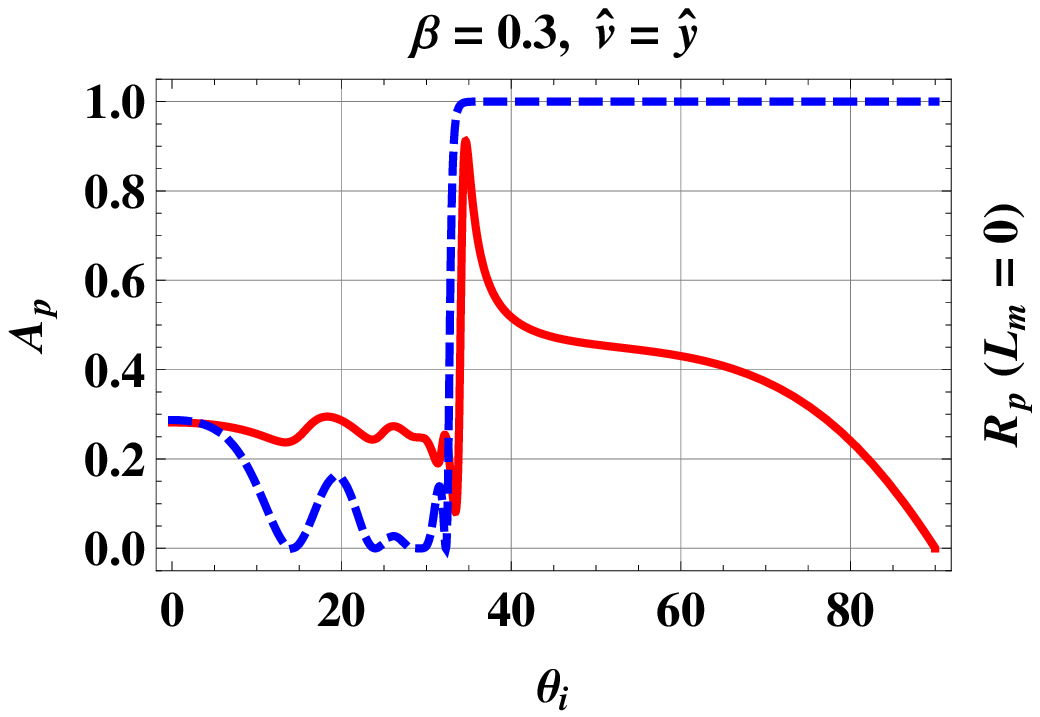} \hfill
\includegraphics[width=3.0in]{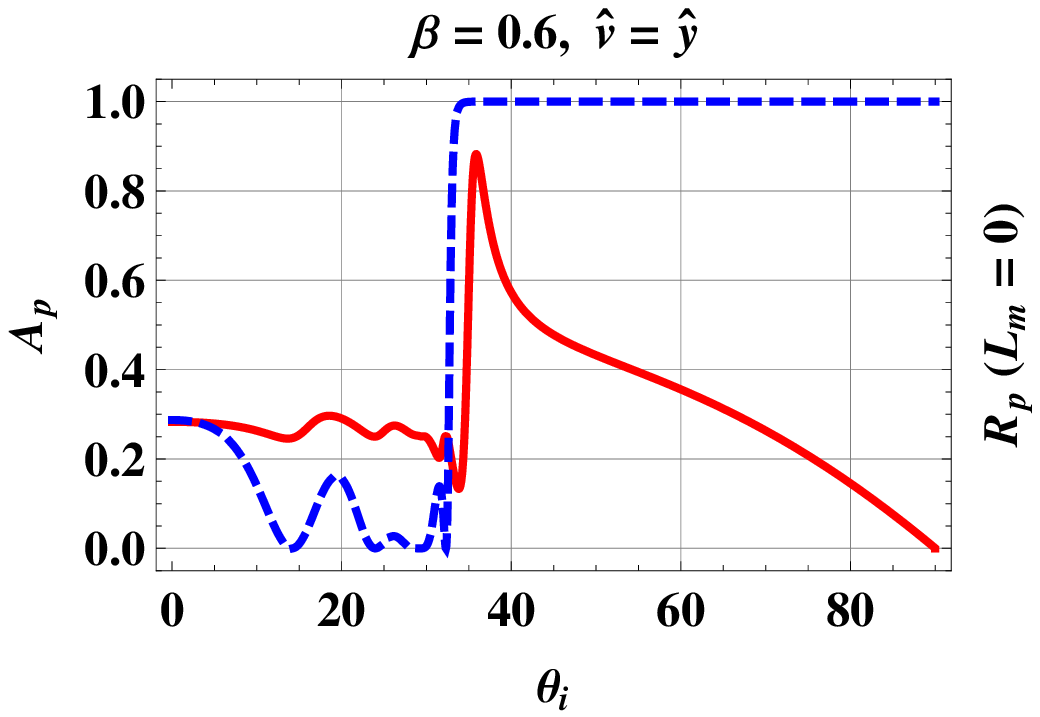}\vspace{10mm} \\
\includegraphics[width=3.0in]{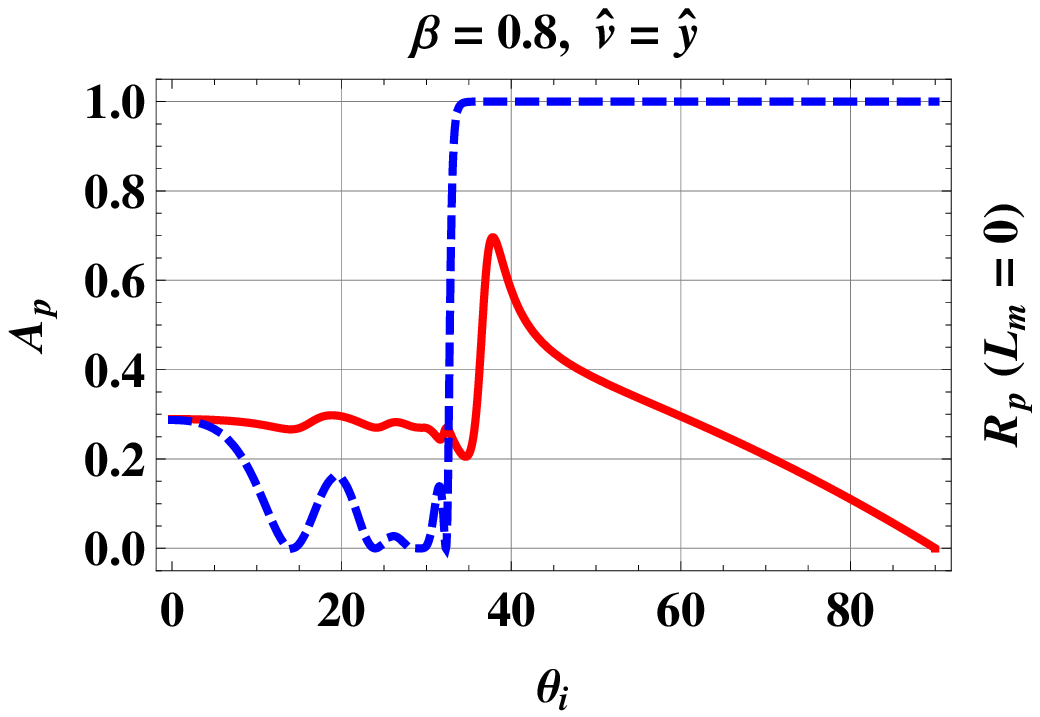} \hfill
\includegraphics[width=3.0in]{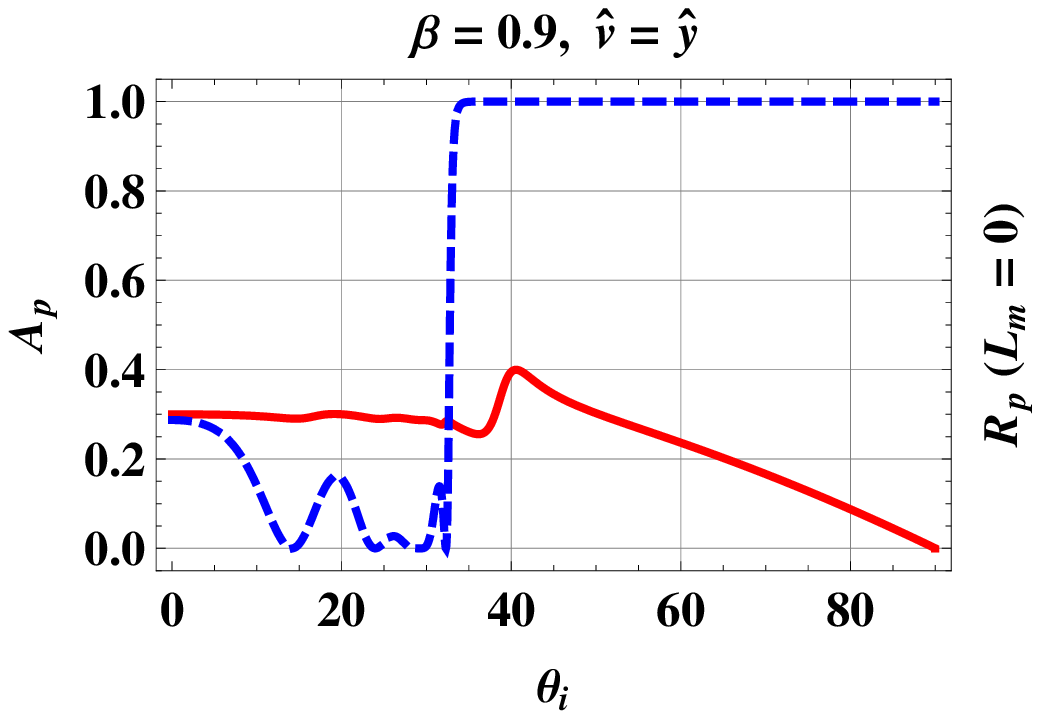}
 \caption{\l{fig8}
 Absorbance $A_p = 1 - (R_{pp} + R_{sp} + T_{pp} + T_{sp})$ (red,
 solid curve) plotted
 versus  angle of incidence $\theta_i$ (in degree) for the scenario
where the metal film  moves in the direction
 $\hat{\#v} = \hat{\#y}$ at relative speeds $\beta \in \lec 0.3, 0.6, 0.8,  0.9 \ric$. Also plotted is the quantity $R_p =
 R_{pp} + R_{sp}$ (blue, dashed curve), calculated when $L_m =0$.
 }
\end{figure}

\newpage

\begin{figure}[!ht]
\centering
\includegraphics[width=3.0in]{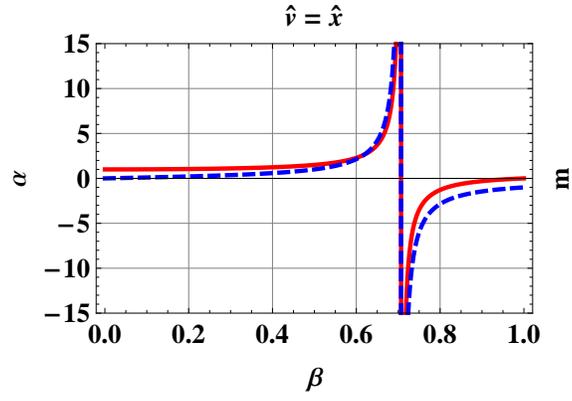}
 \caption{\l{fig9}
 The Minkowski constitutive scalars $\alpha$ (red, solid curve) and
 $ m$ (blue, dashed curve) of the dielectric slab plotted versus relative speed $\beta$ for
 the case where the direction of motion is parallel to the plane of
 incidence. }
\end{figure}

\end{document}